\documentclass{vgtc}                          
\let\ifpdf\relax

\usepackage{mathptmx}
\usepackage{graphicx}
\usepackage{times}
\usepackage{enumitem} 

\vgtccategory{Research}

\vgtcinsertpkg

\title{Interactive Movie Recommendation Through Latent Semantic Analysis and Storytelling}

\author{Kodzo Wegba$^{1}$, Aidong Lu$^{1}$, Yuemeng Li$^{1}$, and Wencheng Wang$^{2}$\\
$^1$ University of North Carolina at Charlotte, USA, \{kwegba1, aidong.lu, yli19\}@uncc.edu\\
$^2$ Chinese Academy of Sciences, China, whn@ios.ac.cn
}

\abstract{
Recommendation has become one of the most important components of online services for improving sale records, however visualization work for online recommendation is still very limited.
This paper presents an interactive recommendation approach with the following two components.
First, rating records are the most widely used data for online recommendation, but they are often processed in high-dimensional spaces that can not be easily understood or interacted with. 
We propose a Latent Semantic Model (LSM) that captures the statistical features of semantic concepts on 2D domains and abstracts user preferences for personal recommendation.
Second, we propose an interactive recommendation approach through a storytelling mechanism for promoting the communication between the user and the recommendation system.
Our approach emphasizes interactivity, explicit user input, and semantic information convey; thus it can be used by general users without any knowledge of recommendation or visualization algorithms.
We validate our model with data statistics and demonstrate our approach with case studies from the MovieLens100K dataset. 
Our approaches of latent semantic analysis and interactive recommendation can also be extended to other network-based visualization applications, including various online recommendation systems.
} 
\teaser{
\centering
\includegraphics[scale=0.65]{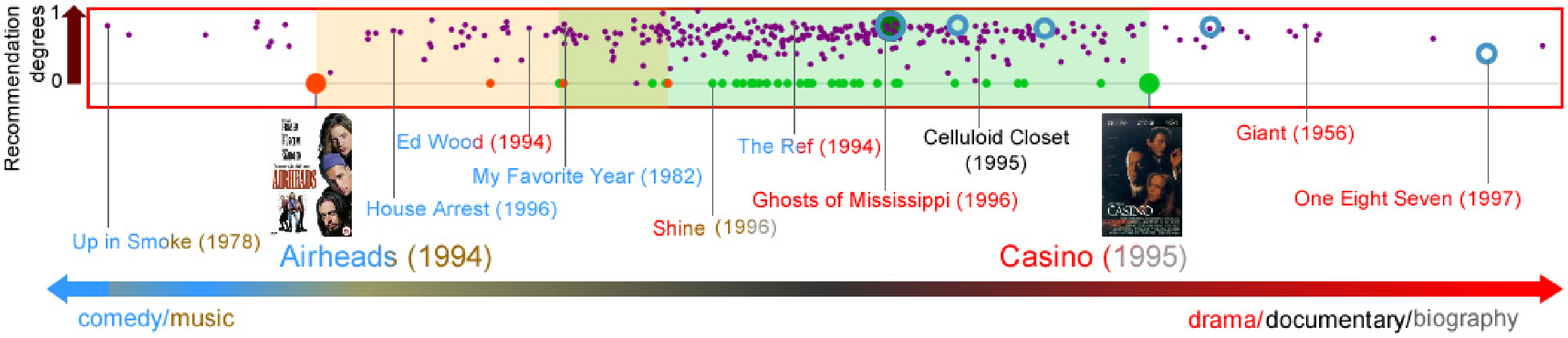}
\caption{
An example from our latent semantic model for interactive recommendation and abstraction of user preferences.
Our approach identifies a 2D visualization domain, where the horizontal axis layouts recommendable movies on a latent dimension between two combined movie features that are selected based on the user's watch history, and the vertical axis uses recommendation degrees to move highly recommendable movies to the top.
This example demonstrates the preference of a user on drama/documentary/biography movies (green zone toward the right) over comedy/music genres (orange zone toward the left).
The movies selected to recommend are enlarged as blue circles, recommendable movies are shown as purple nodes, watched and liked movies as green nodes, and disliked movies as orange nodes.
Two example movie posters, one liked movie ``Casino" and one disliked movie ``Airheads", are also provided to demonstrate the latent dimension.
For illustration purpose, we also add the arrowed line at the bottom and several movie titles to confirm the movie distributions on the visualization domain.
}
\label{svdSemantic}
}

\begin{document}


\firstsection{Introduction}

\maketitle



Recommendation systems have been reported as key pillars of online services~\cite{Gomez-Uribe:2015:NRS:2869770.2843948} for significantly improving sale records~\cite{1167344}.
Nowadays, online retailers and content providers offer a huge amount of products or services, which are often overwhelming for consumers.
To improve user satisfaction and loyalty, Internet leaders like Amazon, Google, and Yahoo are all using recommendation systems to provide personalized suggestions.
However, matching consumers with the most appropriate products is not trivial.
While many recommendation algorithms have been developed, including 
the prevailing top-N recommendation approaches~\cite{Deshpande:2004:ITN:963770.963776}, effective interaction mechanisms for consumers to adjust search preferences or recommendation results are still lacking~\cite{loepp2014choice}.

The challenges for improving user satisfaction of interactive recommendation systems come from several aspects.
First, the satisfaction of consumers vary obviously with other factors such as emotions and situations, which require explicit input from users.
Second, useful information for assisting users to find the right movies, such as the similarities between recommended and rated movies, is often completely hidden from users.
Third, complex recommendation algorithms and visualization systems are often too complicated for general consumers to use.
To the best of our knowledge, there are no previous work on addressing all above challenges.


In this work, we present an interactive recommendation approach that simulates the scenario of an expert recommending movies to a user -- an expert generally selects movies for a user based on his or her watch history, makes recommendations, listens to the feedback, and continues to recommend more movies until the user finds a movie to watch.
Similarly, our approach adopts such a continuous and interactive recommendation process that allows the user to orientate search results actively and explicitly.
For efficiency, we recommend movies in groups, with each group 
containing movies selected from combined movie features that are extracted from watch history.
As shown in Figure~\ref{svdSemantic}, for a user who likes a number of drama, documentary, and biography movies, we visualize the movie distributions on the dimension between the liked and disliked movie features, and make recommendations based on important criteria such as user preference and variety.
This requires us to study the personal movie features from rating statistics and present recommended movies with both useful information and connections with user-rated movies in a succinct, user-friendly manner.

Our work consists of two components, a model of latent semantic analysis and an approach of interactive recommendation.
We first present a latent semantic model (LSM) built upon the high-dimensional latent space factorized from the movie rating records, as they are the most commonly used data for online recommendation.
Our model collects a set of semantic concepts, including like, dislike, familiarity, diversity, typicality, and un-typicality, that can be used for two purposes:
one is to help a user to correlate new recommended movies to the movies the user has watched, the other is to allow the user to specify the preferences of recommendation results explicitly.
We abstract the statistical features of semantic concepts among the high-dimensional latent space and automatically identify suitable 2D domains for interactive visualization.
We validate LSM with a real-life dataset to show that our model is applicable to various users for recommendation.

We then present an interactive recommendation system that recommend movies in a storytelling style to promote the communication between the user and recommendation system.
We employ the LSM to generate recommendation stories by assigning movies with suitable characters, roles, and narrative structures for describing a group of recommended movies.
The recommendation stories are constructed automatically to attract user attention, supported with multi-level visualization and animation effects, and can be adjusted flexibly during the interactive recommendation process.
Different from previous online recommendation systems, our approach reveals information that is generally hidden from users and allows interactive exploration of movie similarities.
We also provide several case studies to demonstrate the usage of LSM and interactive recommendation on different recommendation tasks.

The remainder of this paper is organized as follows. 
We start with related work in Section 2.
We then describe our LSM in Section 3 and interactive recommendation approach in Section 4.
We present the recommendation system in Section 5 and results in Section 6.
Section 7 concludes the work and presents future work.

\section{Related Work}

This section presents the previous works on interactive recommendation, storytelling and narrative visualization approaches.

\subsection{Visualization for Recommendation}

Different from the topic of visualization recommendation that suggests suitable visualization formats given data or tasks, we focus on visualization approaches for recommender systems, where graph visualizations are often adopted.
For example, 
Luo et al.~\cite{luo2009personalized} used hyperbolic and multi-modal view to visualize a recommendation list. 
Kermarrec et al.~\cite{kermarrec2012data} used SVD-like matrix factorization and PCA for global mapping of movie ratings from high dimensions to a two-dimensional space. 
Crnovrsanin et al.~\cite{crnovrsanin2011visual} proposed a 
task-based and information-based network representation for users to 
interact and visualize a recommendation list. 
Vlachos et al.~\cite{vlachos2012recommendation} used bipartite graphs and minimum spanning trees to explore and visualize recommendation results of a movie-actor dataset. 

Interactive recommendation approaches have also been developed.
Gretarsson et al.~\cite{gretarsson2010smallworlds} visualized recommended users in social networks with node-link diagram and grouped relevant nodes on 
the recommendation list in parallel layers. 
Loepp et al. presented an interactive recommendation approach by having users to choose between two sets of sample items iteratively and extracting latent factors~\cite{loepp2014choice}.
Recently, Loepp et al. ~\cite{loepp2015blended} presented MyMovieMixer for interactively expressing user preferences over the hybrid recommendation process.

Our approach dynamically recommends suitable movies to users with the support of interactive storytelling methods.
A key feature of semantic storytelling is that users do not need to understand complex recommendation algorithms or visualization techniques.

\subsection{Storytelling and Narrative Visualization} 

``Storytelling" has a long history and it has become a visualization technique~\cite{Gershon:2001:SIV:381641.381653, 1626183, 6111347, 6412677, 7274435, wojtkowski2002storytelling}. 
While the term of narrative visualization is relatively new~\cite{segel2010narrative}, it also refers to using data stories to improve visual communication~\cite{hullman2011visualization, Hullman:2013:DUS:2553699.2553753, Satyanarayan:2014:ANV:2771495.2771532}.

We focus on techniques of narrative structure, which is a key concept in storytelling and narrative visualization.
It refers to ``a series of events, facts, etc., given in order and with the establishing of connections between them" from the Oxford English Dictionary and it is often simplified to structures like beginning, middle, and end in visualization systems~\cite{segel2010narrative}.
Studies from journalism~\cite{segel2010narrative}
and political messaging and decision-making~\cite{hullman2011visualization} have been performed to understand useful narrative structures for visualization.

Several interactive or automatic storytelling approaches have been developed for applications ranging from general visualization process~\cite{cruz2011generative, 4677364} 
to specific domains. 
For example, Wohlfart and Hauser~\cite{Wohlfart:2007:STP:2384179.2384194}
used storytelling as a guided and interactive visualization presentation approach for medical visualization.
Eccles et al.~\cite{4388992} detected geo-temporal patterns and integrates story narration to increase analytic sense-making cohesion in GeoTime.
Yu et al.~\cite{CGF:CGF1816} generated automatic animations with narrative structures extracted from event graphs for time-varying scientific visualization.
Hullman et al.~\cite{Hullman:2013:DUS:2553699.2553753} presented a graph-driven approach for automatically identifying effective sequences in a set of visualizations to be presented linearly.
Lee et al.~\cite{Lee:2013:STM:2553699.2553755} presented a storytelling process with steps involved in finding insights, turning insights into a narrative, and communicating to an audience. 
Satyanarayan and Heer~\cite{Satyanarayan:2014:ANV:2771495.2771532} developed a model of storytelling abstractions and instantiate the model in Ellipsis with a graphical interface for story authoring.
Wang et al. presented a narrative visualization system that presents literature review as interactive slides with three levels of narrative structures~\cite{Wang:2016:GTL:2968220.2968242}.
Bryan et al.~\cite{7539294} generated textual annotations with a temporal layout and comic strip-style data snapshots for visualizing multidimensional and time-varying data.



Storytelling has been shown to be effective on conveying data in a number of applications~\cite{dragicevic2011temporal, spaulding2013design}.
For visualization tasks, while storytelling did not seem to increase user-engagement in exploration~\cite{boy2015storytelling}, annotated visualizations were proven to be better in balancing graphical salience and relevance~\cite{Hullman:2013:CAG:2470654.2481374} and graph comics were useful to help a general audience understand complex temporal changes quickly~\cite{bach2016telling}.
Nonetheless, it is clear that a flexible creation process should be provided and  adjusted according to application requirements~\cite{mitchell2011limits, 6902874, amini2015understanding}.

Different from other storytelling techniques, we present a highly interactive storytelling approach that simulates human communication with two features - continuous updating stories with or without user inputs and allowing interaction in all stages of exploring data, making a story, and telling a story.

\section{Latent Semantic Model for Interactive Recommendation and Movie Exploration} 
\label{latent}

Our interactive recommendation approach for general users is consisted of two components: LSM for abstracting personal movie preferences (Section 3) and interactive recommendation with storytelling (Section 4).
In this section, we start by introducing the latent space from collaborative filtering algorithms.
We then describe our recommendation approach and measurement of recommendation degrees.
At the end, we present the LSM, which transforms high-dimensional data statistics into recommendation domains according to a set of semantic concepts. 
The LSM is also used to design the recommendation storytelling and user interaction in Section 4.
During this work, the designs of LSM and interactive recommendation system are simultaneously proceeded to ensure that the same set of semantic concepts can be used for both interactive recommendation and exploration of movies.

\subsection{Latent Space from Collaborative Filtering}

To provide an effective interactive recommendation system, we need to integrate a recommendation algorithm into the visualization mechanism.
The latent factor models are the primary approaches of Collaborative Filtering (CF) techniques, which have been successfully adopted by a number of commercial systems~\cite{Koren:2008:FMN:1401890.1401944}.
The latent factor models based on Singular Value Decomposition (SVD) establish recommendations by transforming both movies and users to the same latent factor space, thus making them directly comparable. 
We choose this latent space as it can be used to explore not only recommended movies, but also the distribution patterns of movies and users from the aspect of semantic analysis.

The latent space can be used to interpret a number of preference / relevance features. 
For example, a dimension from comedy to drama can be used to represent the taste of a user favoring these two movie genres.  
In majority of cases, the latent space captures statistical distribution of rating records that combine features from all users, which are often hard to describe or understand directly.
To distinguish users from movies, we reserve special indexing letters of $u$ and $v$ for users and $i$ and $j$ for movies. 
A rating $r_{ui}$ indicates the preference of a user $u$ on a movie $i$, where rating values are in the set of $\{1, 2, ..., 5\}$ with $1$ for no interests to $5$ for strong interests. 

We generate the latent space with the factorization of user-movie rating matrix using SVD. 
For a user-movie matrix $M$ with $m$ users and $n$ movies, the SVD algorithm factorizes $M$ into three matrices such that $M = USV^T$.
It is common to truncate these matrices to yield $U_k$, $S_k$, and $V_k$, in order to decrease the dimensionality of the vector space, and only leave the strongest effects in the model by dropping dimensions with small singular values~\cite{Ekstrand:2011:CFR:2185827.2185828}.
Specifically, the rows of the $U_k$ are the interests of users in each of the $k$ inferred features, and the columns of the matrix $V_k$ are the relevance of movies for each feature. 
The diagonal matrix $S_k$ contains the $k$ biggest singular values of $M$, which are the weights for the preferences, representing the influence of a particular topic on user-movie preferences.

\subsection{Recommendable Movies and Degrees}


In a recommendation approach, we are interested in two things the most: recommendable movies and recommendation degrees for interpreting how likely we would recommend a movie.
We follow a typical recommendation algorithm by adjusting rating history with the normalization of global effects. 
This step balances the tendencies that some users like to give higher ratings than others and some movies receive higher ratings than others, therefore the adjusted ratings $\hat{r_{ui}}$ are more accurate to compare different users or different movies.
Denote $a_{u}$ as the average rating given by user $u$, $a_{i}$ as the average rating of movie $i$, and $A$ and $B$ as the averages of all users and all movies respectively. 
\begin{equation}
\hat{r_{ui}} = r_{ui} - (a_{u} - A) - (a_{i} - B)
\end{equation}

Our recommendation algorithm starts by mapping each user into the latent space by multiplying the adjacency matrix $M$ and the $V_k$ component of SVD.
We denote this product as $C=M \times V_k$, where $C_u$ represents a row vector for user $u$ from the matrix $C$.
We then compute the cosine similarity $s_{uv}$ between users $u$ and $v$ 
with the $C_u$ and $C_v$ coordinates~\cite{leskovec2014mining}.
\begin{equation}
s_{uv} =  \frac{\displaystyle\sum_{l=1}^{k} {C_u}_l \times {C_v}_l}
  			 {\sqrt{\displaystyle\sum_{l=1}^{k} ({C_u}_l)^2 } \times 
  			  \sqrt{\displaystyle\sum_{l=1}^{k} ({C_v}_l)^2 } }    
\label{similarity}     
\end{equation}

Next, we select a list of similar users with positive similarity coefficients, as recommendable movies are generally selected based on ratings from similar users.
We denote this set as $S_u=\lbrace v | s_{uv} \geq 0 \rbrace $ for user $u$.
For a new user with no rating record, all the similarity coefficients $s_{uv}$ are zero and the set $S_u$ contains all the users.

We further select the list of recommendable movies $L_u$ for the user $u$ by choosing movies that have not been watched by $u$, but received positive ratings from similar users as follows, where the positive rating threshold $w_c$ is initialized as $3$ and can be adjusted for different numbers of recommendable movies.
\begin{equation}
L_u=\lbrace i | v \in S_u, \hat{r_{vi}} \ge w_c \ and \ r_{ui}=0 \rbrace. 
\end{equation}

In addition, we measure the recommendation degree of a movie $i$ for user $u$ by $b_{ui}$.
The movies with high $b_{ui}$ degrees are more likely to be recommended during the interactive recommendation process.
It is calculated as the average rating from similar users with positive cosine similarities. 
For any $i \in L_u$,
\begin{equation}
b_{ui} = \sum_{v \in S_u \ and \  r_{vi} \ne 0} \{r_{vi} \times 
s_{uv}\} / \sum_{v \in S_u \ and \ r_{vi} \ne 0} \{r_{vi}\}
\label{ratingScore}
\end{equation}
We normalize the $s_{uv}$ and $b_{ui}$ values to range of $[-1, 1]$ and $[0,1]$ respectively to balance the differences among users.


\begin{figure}
\centering
\includegraphics[width = 3.4in]{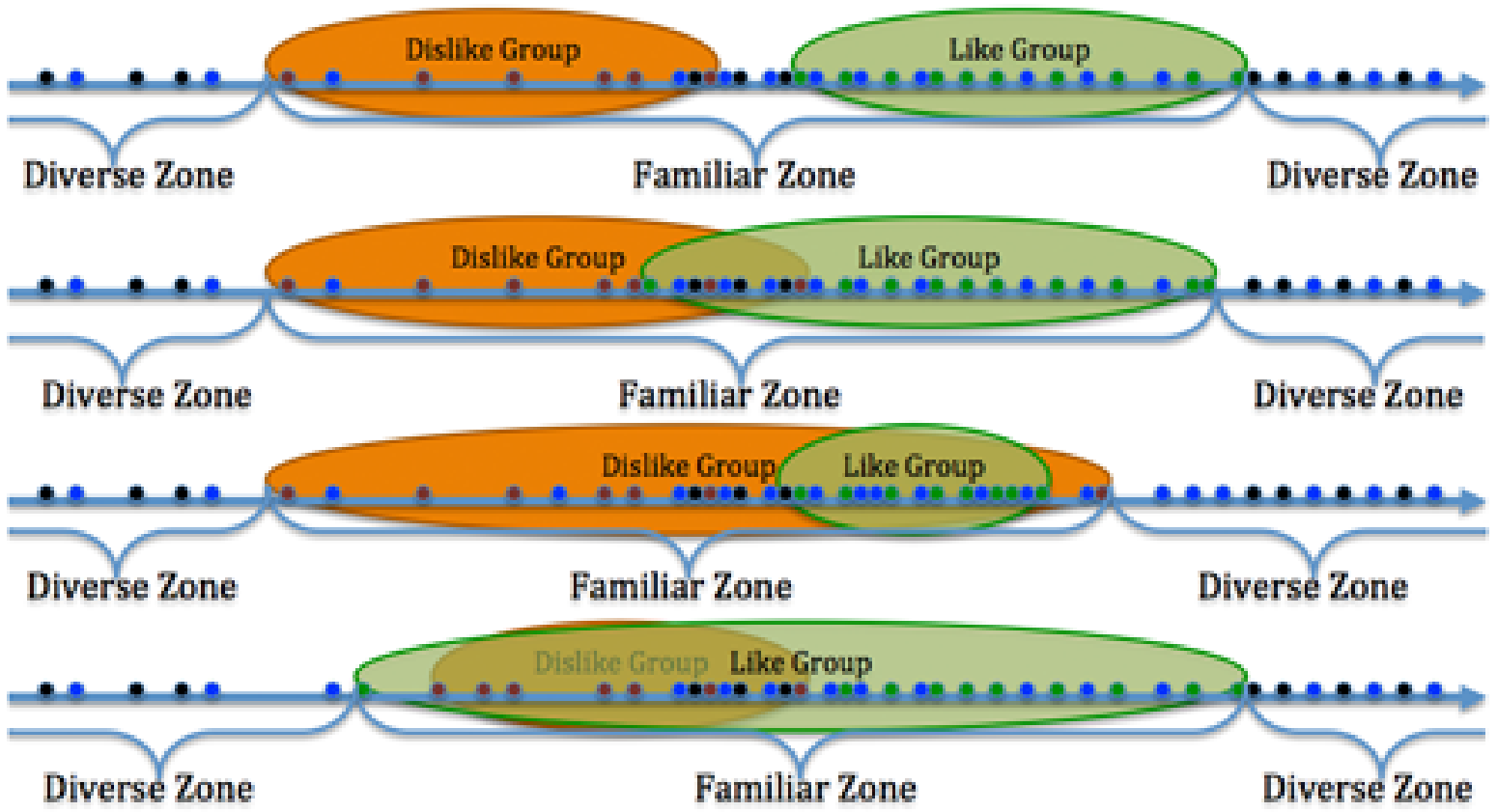}\\
LSM mode (a) -- Variations based on the Like/Dislike groups\\
\includegraphics[width = 3.4in]{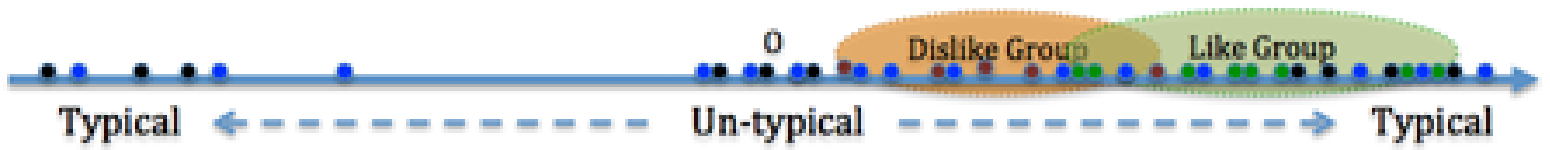}\\
LSM mode (b)
\caption{The LSM maps the distribution of a set of semantic concepts in the latent space.
We separate the two modes of LSM for clarity -- (a) includes concepts of like, dislike, familiarity and diversity and (b) includes concepts of typicality and un-typicality.
The movie nodes are colored based on the groups -- liked in green, disliked in orange, recommendable in blue, and not recommendable in black.
}
\label{svd}
\end{figure}

\subsection{Latent Semantic Model}

While the latent space has been widely used in recommendation algorithms, it is a high-dimensional space and not directly suitable for visualization or interaction. 
Our goal of the LSM is to identify semantic concepts and visualization domains that can be used in interactive recommendation and exploration of movie similarities.
Our approach is consisted of the following three steps: separating movie groups, identifying a set of semantic concepts, and selecting suitable latent dimensions for interactive recommendation.

\subsubsection{Separating Movie Groups}

To identify recommendable movies, we separate all the movies in the database to five groups based on the rating history of the user and our estimation of recommendation degrees.
For any user $u$, each movie belongs to one and only one of the following groups.

\begin{itemize}
\vspace{-2mm}
\item Like (with positive ratings from $u$): $G_+ = \{i | {r_{ui}} \geq \tau_{+}\}$
\vspace{-2mm}
\item Dislike (with negative ratings from $u$): $G_- = \{i | 0 < {r_{ui}} < \tau_{-}\}$
\vspace{-2mm}
\item Neutral: $G_{neu} = \{i | \tau_{-} \leq r_{ui} <\tau_{+}\}$
\vspace{-2mm}
\item Recommendable (with positive recommendation degrees):\\ $G_r = \{i | b_{ui} \geq \tau_{r}\}$ and not specified as ``thumb-down" by the user during interaction recommendation process
\vspace{-2mm}
\item Not recommendable (with negative recommendation degrees): $G_n = \{i | b_{ui} < \tau_{r}\}$ or specified as ``thumb-down" by the user during interaction recommendation process
\end{itemize}

The thresholds of positive rating $\tau_{+}$ and negative rating $\tau_{-}$ are set to value $3$, and the threshold of recommendation degree $\tau_{r}$ to value $0$ initially.
The values can be adjusted to control the number of movies in each group.

\subsubsection{Semantic Concepts for Recommendation}

In everyday life, we often describe an object with several terms, such as families, friends, and enemies for a person.
Similarly, a set of semantic concepts can be used to describe movies for a quick impression, which may assist users to find movies effectively.

In addition to ``like" and ``dislike", we have identified the following four semantic concepts for recommendation based on two criteria -- concepts that are often used in recommendation for the general public; 
and concepts with clear distribution features in the latent space.

\begin{itemize}
\item Familiarity -- movie styles that the user has already watched;
\vspace{-2mm}
\item Diversity -- movie styles that the user is not familiar with;
\vspace{-2mm}
\item Typicality -- movie styles that can be well defined based on combined movie features;
\vspace{-2mm}
\item Un-typicality -- movie styles that are unclear to a feature.
\end{itemize}

\begin{figure}
\centering
\includegraphics[width = 3.4in]{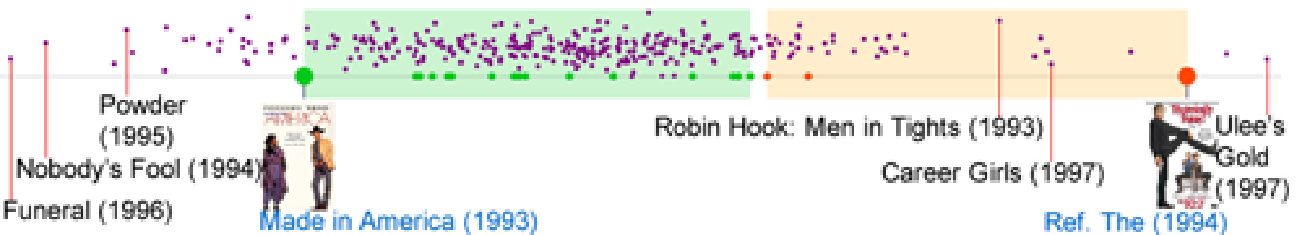}\\
\vspace{+1mm}
\includegraphics[width = 3.4in]{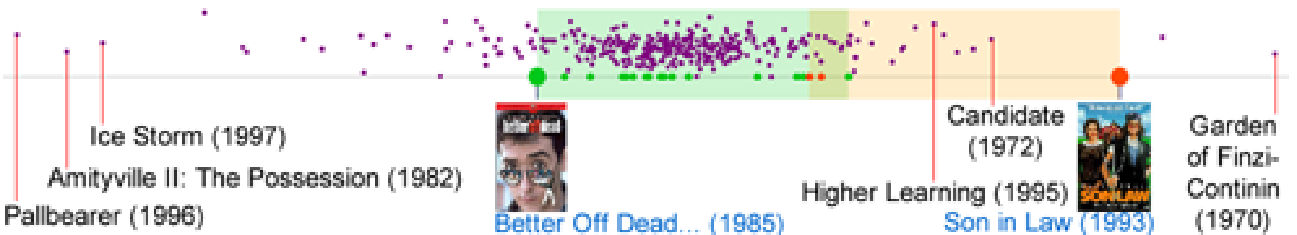}\\
\vspace{+1mm}
Good latent dimensions for visualization and interaction\\
\vspace{+1mm}
\includegraphics[width = 3.4in]{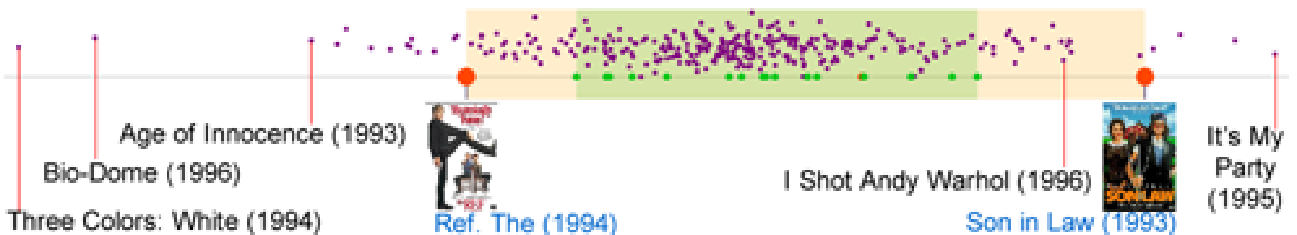}\\
\vspace{+1mm}
\includegraphics[width = 3.4in]{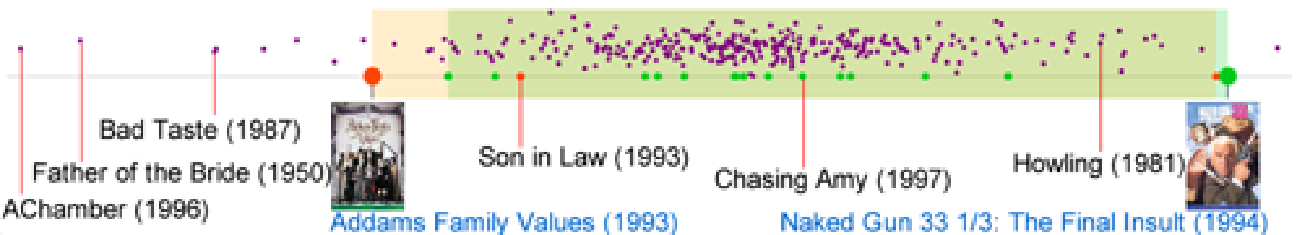}\\
\vspace{+1mm}
Bad latent dimensions as visualization and interaction
\caption{Good and bad example latent dimensions that are selected based on criteria described in Section 3.3.3.
The liked and disliked zones are highlighted with shaded rectangles in green and orange respectively.
The recommendable movies are colored in purple, ``liked" movies in green, and ``disliked" movies in orange.
}
\label{svdexample}
\end{figure}

Next, we describe the distribution features of semantic concepts in the latent space.
As illustrated in Figure~\ref{svd}, we can identify the familiar zone by including both the like and dislike groups -- all the movies that have been watched by the user $u$.
The diverse zones are regions outside the familiar zone and there are generally two diverse zones on each side of a latent dimension.
Figure~\ref{svdexample} uses examples from real scenarios to show the variations of the LSM mode (a).

The LSM mode (b) between typicality and un-typicality utilizes the distance of a movie to the Origin of the latent space.
Since both the $U_k$ and $V_k$ from the latent space demonstrate clustering features of similar users and movies~\cite{Pu:2013:UIR:2507157.2507178}, the distances in the latent space can interpret the similarity degrees.
For example, on a latent dimension which contains comedy movies on one side and drama movies on the other, the movies with combined comedy/drama or other genres are distributed near the Origin.
This feature is similar to spectral spaces that nodes with few connections are often located close to the Origin~\cite{6231611}.
What's important is that this feature preserves for general latent dimensions describing combined movie features or styles, with only un-typical movies corresponding to the latent dimension located close to the Origin.
As shown in Figure~\ref{svd} mode (b), any latent dimension can be modeled as typical to untypical to typical zones.
The two typical zones correspond to two opposite features represented by the latent dimension.

While the distribution features of semantic concepts on the latent dimensions are clear, the relative locations among the semantic zones 
vary depending on the Origin and the familiar zone.
For example, the Like and Dislike zones can appear on either side of the Origin or Overlap with the Origin.
Nonetheless, they all provide valuable semantic information for interactive recommendation.
We generally adjust thresholds of $\tau_{+}$, $\tau_{-}$, and $\tau_{r}$ to make sure that all the semantic zones contain recommendable movies.

\subsubsection{Selecting Dimensions for Visualization and Interaction}

In the latent space, each dimension describes certain joint statistical features of movies and users.
We are interested in searching for dimensions that can help users understand movie relationships and are suitable for visualization and interaction.
We need to select not only one, but also a set of suitable dimensions, so that multiple aspects of recommendable movies can be covered.
However, not all the dimensions in the $k$-dimensional latent space are suitable for recommendation, as they may represent features of irrelevant movies or users, repeated features from other dimensions, or features that are hard to understand.
Therefore, we go through the following process of selection based upon the distribution of the semantic concepts.

The selection is based on the combined information of user history, recommendation degrees, and explicit information from user interaction -- thumb up/down for specific movies, which is described in the section 4.
We prefer to choose the dimensions that separate the semantic zones with the following three factors:
\begin{description}[style=unboxed,leftmargin=0cm]
\vspace{-2mm}
\item[Group sizes.] 
For each dimension, we measure the region of the like group as $R_+$, the dislike group as $R_-$, the overlapped region as $R_{o}$, and the combined range as $R$.
It is ideal that $R_+$ occupies a large portion of $R$, as majority recommendable movies are selected within or close to this region.
We also prefer that $R_+$ and $R_-$ do not cover the entire dimension, so that there is room for diverse zones.
This factor is simplified as $R_+ / R$.
\vspace{-2mm}
\item[The overlapping region.]
We try to avoid the third and fourth cases of LSM mode (a) in Figure~\ref{svd}, which happens when one group is located inside another.
They are less desirable as the meanings of such dimensions are hard to describe and understand.
We set large penalty to avoid large overlapping ratios for $R_o / R_+$ and $R_o / R_-$. 
\vspace{-2mm}
\item[The distribution of recommendable movies.]
Inside $R_+$, it is ideal that the recommendable movies are evenly distributed. 
This factor helps to remove the dimensions with many movies mapping to a small range, which indicates that these movie features and differences are not well represented on the dimensions.
We measure this factor with standard deviation $\theta_+$ of recommendable movies in $R_+$.
\end{description}

Specifically, we use the following $D_v$ equation to measure if a latent dimension $v$ is suitable as a visualization domain, where $w_+$, $w_{o}$ and $w_{\theta}$ are the weights for the three factors described above.
\begin{equation}
D_v =  (\frac{R_+}{R})^{w_+} \times (1 - \frac{R_o}{R_+})^{w_o}  \times (1 - \frac{R_o}{R_-})^{w_o} \times ( \frac{R_+}{\theta_+})^{w_{\theta}} 
\label{equ:dv}
\end{equation}

Next, we filter the set $S_{d} = \{v | D_v > \tau_{v} \}$, composed of latent dimensions with high $D_v$ values, by removing similar dimensions.
This is achieved by comparing the locations of movies to the groups on the two dimensions.
Among the similar dimensions with high $D_s$ values, only the dimension with the highest $D_v$ value is kept in $S_{d}$.
\begin{equation}
D_s(p, q) = \sum_{i \in L_u} w_i \times 
[(i \in R_{+p}) \bigoplus (i \in R_{+q}) + (i \in R_{-p}) \bigoplus (i \in R_{-q})]
\label{equ:ds}
\end{equation}
where, $p \in S_{d}$ and $q \in S_{d}$ are latent dimensions, $w_i$ is the weight for each movie to incorporate user preference, $R_{+p}$ and $R_{-p}$ are the ranges of like and dislike groups on $p$.
We set $w_i = 1$ for all movies initially and double the value when the user clicks the thumb-up/down buttons.

\subsection{Model Validation}

We validate LSM with a real dataset from two aspects.
First, we test LSM on all the users in the MovieLens 100K dataset~\cite{MovieLens100k}.
For all the results in this paper, we use 4 for $\tau_+$, 2 for  $\tau_-$, 5 for $w_+$, and 10 for $w_o$ and $w_{\theta}$.
The best latent dimensions are automatically selected and used to measure how well LSM can distinguish the like and dislike regions.
The results show that the best dimensions for all the users contain ideal group distributions in either case 1 or case 2 of Figure~\ref{svd} (a), indicating that LSM can be applied to users with various rating histories.

Second, we observe the distributions of recommendable movies on the best latent dimensions.
Since every recommendable movie can be in the search results, we collect the total amount of $b_{ui}$ for all movies in a group. 
It is worth mentioning that the dislike region may also contain recommendable movies, as the choices of recommendation come from multiple aspects. 
Also, cases like two users rated one movie similarly while rating another movie very differently can complicate the statistical distributions.
As shown in the table~\ref{tablevalid}, the average in $R_+$ of all users is significantly higher than that of $R_-$.
If we remove the overlapping regions from $R_+$ and $R_-$, the difference is more significant. 
We also compute the Pearson correlation to compare the values from two pairs in statistics.
By removing the overlapping regions, the second pair is very close to no correlation.
This result shows that LSM captures the majority recommendable movies in the like region for making recommendations.

\vspace{-4ex}
\begin{table}[htb]
\caption{Comparison of the sum of $b_{ui}$ in different regions}
\vspace{1ex}
\label{tablevalid}
\centering
\begin{tabular}{| c | c | c | | c | c | c |}
\hline
$\sum_{i \in R_+}$ & $\sum_{i \in R_-}$ & Pearson & $\sum_{i \in R_+ \bigcap \overline{R_o}}$ & $\sum_{i \in R_- \bigcap \overline{R_o}}$ & Pearson \\
\hline
151.0 & 101.7 & 0.54 & 62.5 & 13.1 & -0.05\\
\hline
\end{tabular}
\end{table}
\vspace{-2ex}

\section{Interactive Recommendation through Storytelling}

We start this section by describing how we connect the process of recommendation to storytelling.
Then, we present our strategies to construct recommendation stories automatically through the LSM from the aspects of characters, roles, user interaction and narrative structures respectively.

\subsection{Connecting Recommendation to Storytelling}

As described in the introduction, we propose interactive recommendation to simulate the scenario of an expert recommending movies to a user.
During recommendation, the expert generally presents one or several movies and provides reasons of recommendation, such as high popularity, high similarity to a movie the user likes, or special features.
The user may respond by indicating his or her preferences on the recommended movies, such as ``recommend more movies like this" or ``no more movies like that".
This process continues until the user finds an interesting movie to watch.

\begin{figure}
\centering
\includegraphics[width = 3in]{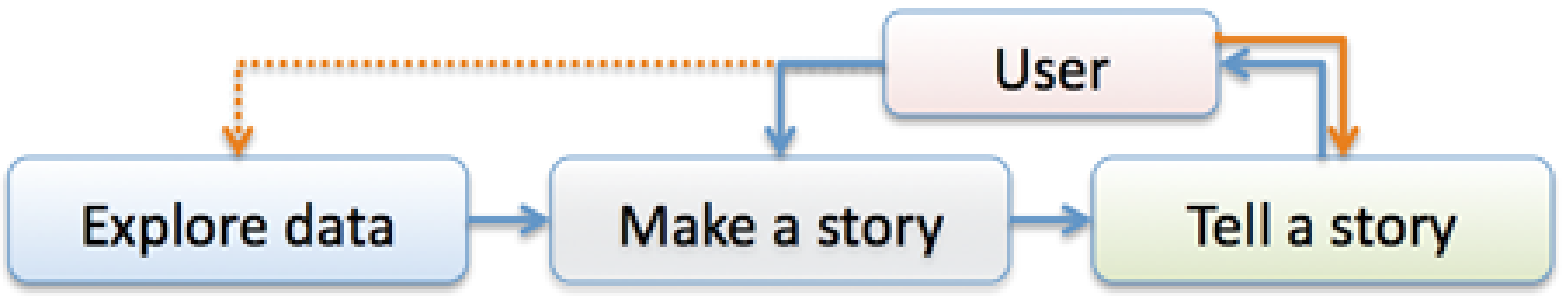}
\caption{Pipeline of storytelling with continuous communication between the user and our interactive recommendation system.
Users can interact with the system at any time to indicate preferences and explore movie information to accelerate recommendation process.
}
\label{IT}
\end{figure}

To simulate this communication process, we present a storytelling mechanism that treats the procedure of recommending movies as telling a story.
We design a \textbf{recommendation story} as a set of recommendable movies and brief reasons of recommendation and an \textbf{interactive recommendation} approach as a continuous storytelling process, which can be adjusted promptly with user interaction.
As shown in Figure~\ref{IT}, the interactive storytelling pipeline starts with exploring movie database for a user, selecting recommendable movies, and collecting necessary information using LSM described in Section 3.
Then, the second step of ``make a story" is to generate a recommendation story automatically with the approach described in this section.
The third step ``tell a story" is to present a story as an animation sequence described in Section 5.

The loop among ``make a story", ``tell a story", and ``user" provides the proposed continuous and interactive process of recommendation until a desirable movie is identified.
The arrow from ``user" to ``make a story" indicates that the user can provide feedback to request new recommendation stories that reflect user inputs.
The arrow from ``user" to ``tell a story" indicates that the user can  adjust the storytelling animation interactively, such as replaying a recommended movie or finishing the story immediately.
If the user does not provide an input, the loop continues to different recommendation stories to achieve the effect of ``how about some other types of movies you may like?"
The automatic switch between different recommendation stories can avoid users getting bored from similar movies.
During the process, the user can also use our interaction tools to explore additional information of movies.

Different from the previous storytelling processes~\cite{7274435} that are mainly an ordered sequence of the three components - explore data, make and tell a story, our interactive recommendation approach is supported by 1) interaction functions that allow a user to interact with the storytelling pipeline at any time during the process, 2) automatic construction of narrative structure that allows new and adjusted recommendation stories to be generated continuously.

\subsection{Character}

The character in an ordinary story is generally a person. 
In our recommendation stories, the characters are movies.
Similar to the human characters, each movie character maintains a different relationship with the user, such as a rated movie, a favorite movie, or a recommendable movie.
The movie characters are also related to each other, such as being rated by the same users or have similar rating averages.
Since our focus is recommendation, the leading characters are recommendable movies, whose relationships with the user can be represented with LSM and recommendation degrees.

\subsection{Role}

A role describes what function each character serves in the story.
The roles of recommendable movies provide a mechanism for users to explore the movie database. 
Before going through the movie details, the roles of a movie provide a quick catch of the movie features, such as a typical drama movie which is very similar to one of the user's favorite movies.
This provides the user a quick way to find several interested movies to explore.

In recommendation stories, we use the semantic concepts from LSM to describe the roles, such as a ``liked" movie that the user rated high or a ``typical" movie that has strong features of certain movie genres.
Each movie character can play multiple roles, such as familiarity and un-typicality, just like a human character can be both a colleague and a friend. 
The actual roles of a movie character are determined by the locations among the semantic zones on selected latent dimensions, as shown in Figure~\ref{svd}.

\subsection{User Interaction During Recommendation}

For interactive recommendation, the user interaction becomes an important component of the storytelling pipeline.
To allow active user interaction with the storytelling pipeline shown in Figure~\ref{IT}, we provide three groups of explicit interaction functions as follows:

The first group of interaction functions is from ``user" to "make a story". 
Corresponding to the roles of a movie, users can specify the preferred movie types between options of familiar ($f$) / diverse and typical ($t$) / untypical movies.
The parameter values become effective immediately on generating the new recommendation stories.
For a specific movie, the user can click thumbs-up (like) and thumbs-down (dislike) buttons, so that the specified movie is moved to the like or dislike groups (and the group of not recommendable, so that it is removed from the recommendation process).
We also increase the $w_i$ value (a parameter to control the effects of user selection) of the movie in equations~\ref{equ:dv} and ~\ref{equ:ds} for choosing latent dimensions for new stories.
For each dimension $v$, the new measurement $D'_v$ contains components from both data distribution by $D_v$ and user interaction. 
We detect if the user preference is aligned with LSM, especially if a liked movie is in the like range and if a disliked movie is in the disliked range.
\begin{equation}
D'_v = D_v  + w_{i} \times (\sum_{i \in (U_+ \cap R_+) \cup (U_- \cap R_-)} {b_{ui}} - \sum_{i \in (U_+ \cap R_-) \cup (U_- \cap R_+)} {b_{ui}})
\end{equation}
where $U_+$ and $U_-$ are the sets containing all liked or disliked movies specified by the user.

The second group of interaction functions is from ``user" to "tell a story". 
To control the animated storytelling, the users can replay, pause, and stop the current recommendation story, or play more stories (the default is continuing to recommend additional movies).

The third group is to explore movie details.
If finding an interesting movie, the user can click the movie poster or movie node on the interface to view details.
The users can also mouse over a movie node anytime to reveal a set of basic information, including user rating, average rating, popularity, title and genres.


\subsection{Automatic Generation of Narrative Structures}

A narrative structure refers to the sequence of events in a story.
In recommendation stories, each event is a recommendable movie and brief reasons of recommendation. 
The sequence of a set of recommendable movies in a narrative structure is crucial to improve users' understanding of the movies effectively.

Considering the short attention of users in online recommendation, we prefer simple stories that can finish in a very short duration. Therefore, we generate each recommendation story only with one latent dimension selected with LSM.
As each latent dimension reflects one combined movie / user feature, such recommendation stories simulate the effect that we recommend movies from one combined movie feature to another, such as popular drama/comedy movies to unpopular documentary movies.
Other designs of recommendation stories using our LSM are also feasible.
For example, long stories can be generated by connecting different latent dimensions. 
Due to the focus of our interactive recommendation approach, we only use short stories in this work.

Based on a latent dimension, we try to maintain smooth story transitions by generating linear narrative structures among the four semantic concepts: familiarity, diversity, typicality, and un-typicality.
This is achieved by identifying a starting point, choosing narrative structures based on user preferences, and selecting recommendable movies.

The \textbf{starting point} of a story is determined according to user preferences of $f$ and $p$.
The default values are $0.5$ for both $f$ and $p$, although we favor typical over un-typical and familiar over diverse movies, which is consistent with the preferences of majority users.
The following list is the order we set as default.

\vspace{+1mm}
\noindent {High familiarity:} starting from the like group

\vspace{+1mm}
\noindent {High diversity:} starting from the diverse zone closer to like group

\vspace{+1mm}
\noindent {High typicality:} starting from the typical zone closer to like group

\vspace{+1mm}
\noindent {High un-typicality:} starting from the un-typical zone
\vspace{+1mm}

\begin{figure}
\centering
\includegraphics[width = 2.5in]{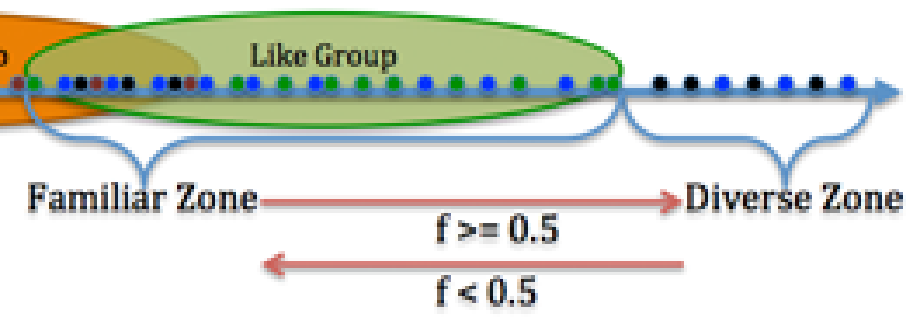}
\includegraphics[width = 2.5in]{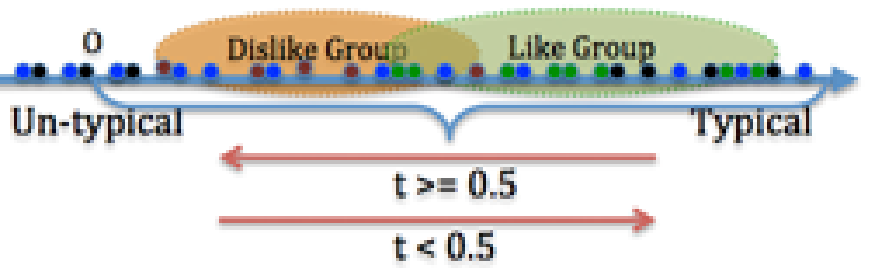}\\
\caption{The four linear narrative structures based on the user preferences of t -- typical, u=1-t -- un-typical, f -- familiar, and d=1-f -- diverse. 
}
\label{path}
\end{figure}

The \textbf{order} of recommended movies also considers user preferences.
As shown in Figure~\ref{path}, we use the four narrative structures
to cover all combinations of the two user preferences of $f$ and $t$.
The narrative structures are designed to be linear sequences, so that users can expect very similar narrative visualization during the interactive recommendation process.
Since the narrative structure is on one latent dimension each time, two zones from the LSM are involved.
The ranges of the latent dimension to select recommended movies are also shown in Figure~\ref{path}.
During the interactive storytelling process, we switch narrative structures between the two options randomly to avoid simply repeated stories.


The \textbf{selection of recommended movies} is based on the sample rates computed according to user preferences of $f$ and $t$.
Assume a set $G_T$ of $T$ recommended movies is selected for each narrative structure.
When the structure is between typicality and un-typicality, we determine the number of recommended movies from the typical zone as $s_t$ and un-typical zone as $s_{u}$:
\begin{equation}
\centering
s_t = \lceil t * T \rceil; \ \ \ \ \ 
s_{u} = T - s_t
\end{equation}
Similarly, when the structure is between familiarity and diversity, the sampling numbers for familiar zone $s_f$ and diverse zone $s_d$ are: 
\begin{equation}
\centering
s_f = \lceil f * T \rceil; \ \ \ \ \ 
s_d = T - s_f
\end{equation}

Inside each zone, we select recommended movies with the following procedure that is composed of a local sampling and a random test procedure.
We first randomly pick a location $l$ and choose the best candidate within a local window by combining all three factors: the recommendation degree $b_{ui}$ of a candidate movie $i$, the distance of movie $i$ to location $l$, and the similarity of movie $i$ to the set of $q$ movies that are specified by the user.
On the latent dimension $V_p$, assume the location of movie $i$ is $V_p(i)$.
The effect of a user specified movie is set to be within a location window $\delta$ with a truncated Gaussian function $G_q()$, with high weights for thumb-up movies and low weights for thumb-down movies.
The movie $i$ with the highest value from the following combined measurement is selected as the best candidate.
\begin{equation}
b_{ui} \times \overbrace{ G_1(| V_p(u_1) - V_p(i) |) \times ... \times G_q(| V_p(u_q) - V_p(i) |)}^{q = | U_+ \cup U_- |} / | V_p(i)- l |
\end{equation}


The purpose of an additional random test is to ensure that our selections of recommendable movies are consistent with both user preferences of $f$ and $t$, although only one factor is used to determine the narrative order.
If the selected movie $i$ passes the random test, we add it to the selected set $G_T$; otherwise we pick another random location and perform the local sampling again.
For example, for the narrative structure between familiarity and diversity, we try to maintain an average typicality value close to the user preference $t$.
We measure the typicality value of a movie $i$ as $t(i) = | V_p(i) |$.
The test is determined by if the movie $i$ can make the average typicality value closer to $t$.
\begin{equation}
| \frac{(\sum_{i \in G_T}{t(i)}) + t(i)} {| G_T | + 1} - t | < | \frac{\sum_{i \in G_T}{t(i)}} {| G_T |} - t |
\end{equation}
Similarly, for the narrative structure between typicality and un-typicality, we measure the familiarity value of a movie $i$ given the center location of the like group $c_+$ as $f(i) = | V_p(i) - c_+ |$.
The random test is performed by replacing the $t(i)$ with $f(i)$ in the above equation.

Since the movie database is generally large, we can assume that there are always enough movies to recommend.
In the cases that the recommendable movies run out, we can adjust the parameters $\tau_+$ and $\tau_-$ to include additional movies.

\section{Interactive Recommendation System}

\begin{figure}[htb]
\centering
\includegraphics[scale=0.28]{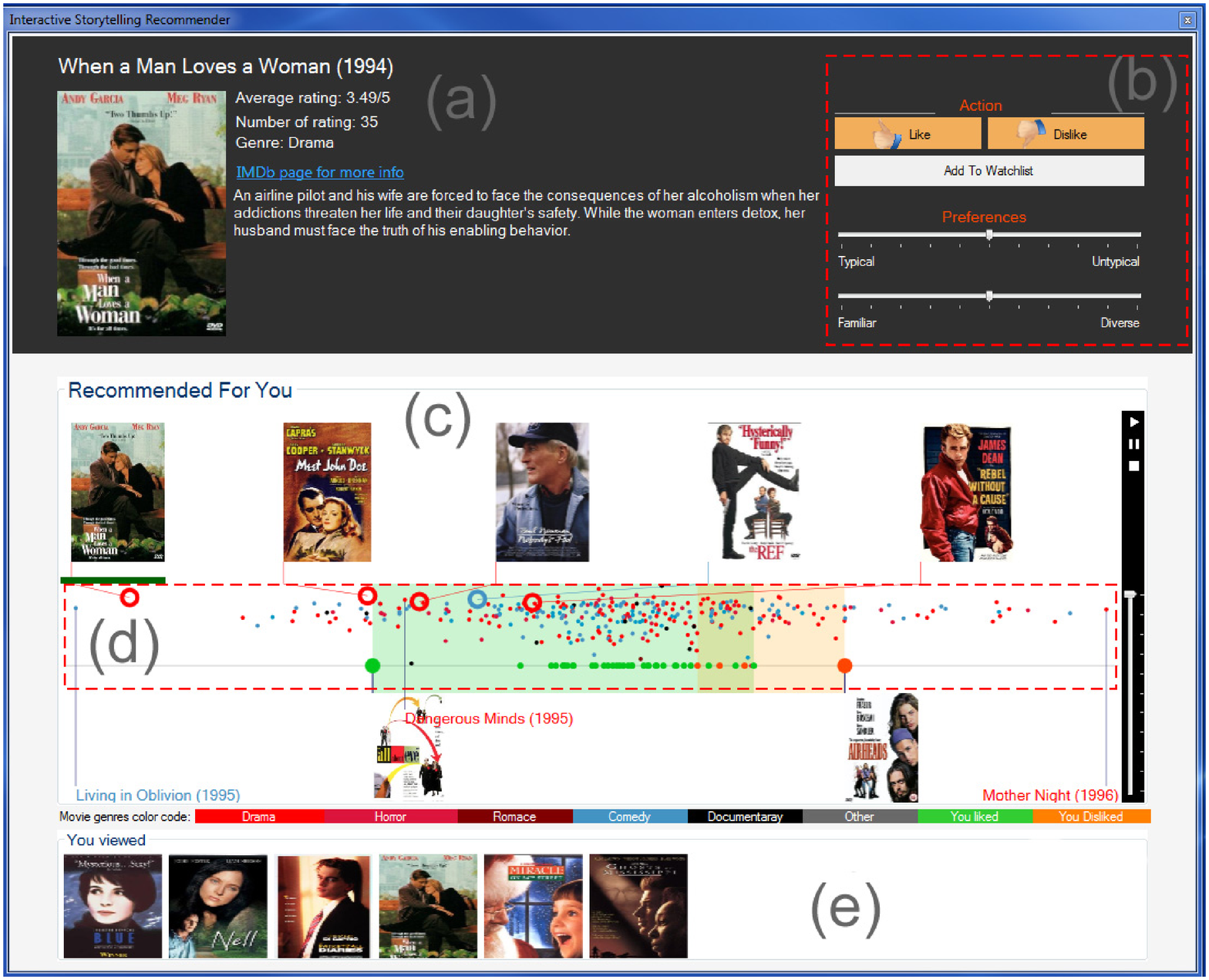}
\caption{The interface of our recommendation system.
Similar to online movie recommendation systems, our interface includes (a) basic information of a selected movie, (c) a list of recommendable movies, and (e) watch history.
We also add (b) for explicit input of user preferences and (d) visualization for exploring recommendable movies.
}
\label{interface}
\end{figure}

The interface of our system is designed to be consistent with popular commercial recommendation systems, such as Youtube, Netflix and Amazon Movie. 
As shown in Figure~\ref{interface}, our system contains three common components -- an enlarged movie poster with information (a), the list of top-N recommended movies (c), and additional information at the bottom (e).
We add a small area for user interaction (b) and a visualization domain (d) to support interactive recommendation and exploration of movie database.
Our system also supports the following features of interactive recommendation.


\subsection{Example-based Narrative Structure}

The visualization of the latent dimension plays an important role in illustrating the recommendable movies and the movie distributions.
To describe the narrative structure visually, we design an example-based approach shown in Figures~\ref{svdSemantic}, \ref{svdexample}, \ref{interface}, which uses two movies that have been watched by the user, one on each side of the familiar zone, to describe the combined features of that latent dimension.
Since the combined features are generated simply from data statistics, they generally cannot be described with languages or equations.
The example movies help the users to understand the overall trend of the latent dimension and provide a quick impression of a new movie by the distances to the two examples.
All the recommendable movies are located on the 2D domain of latent dimension and recommendation degree, providing additional visual cues for recommendation reasons.

\subsection{Multi-Level Visualization}

The reasons to recommend a movie can be multifold. 
While the reasons supported by LSM mainly come from the aspects of statistical similarity and semantic analysis, we can provide a variety of information as the ``brief" reasons of a recommendation, such as a typical drama movie in the familiar zone of a user, which is similar to an example movie shown with the poster.
We organize the information on the visualization domain at three levels, so that the storytelling animation can follow the levels to achieve the effect that additional information is introduced gradually.
Users can stop at any time if they are not interested in the movie and continue to the next recommendation.

\begin{figure*}[htb]
\includegraphics[width = 1.64in]{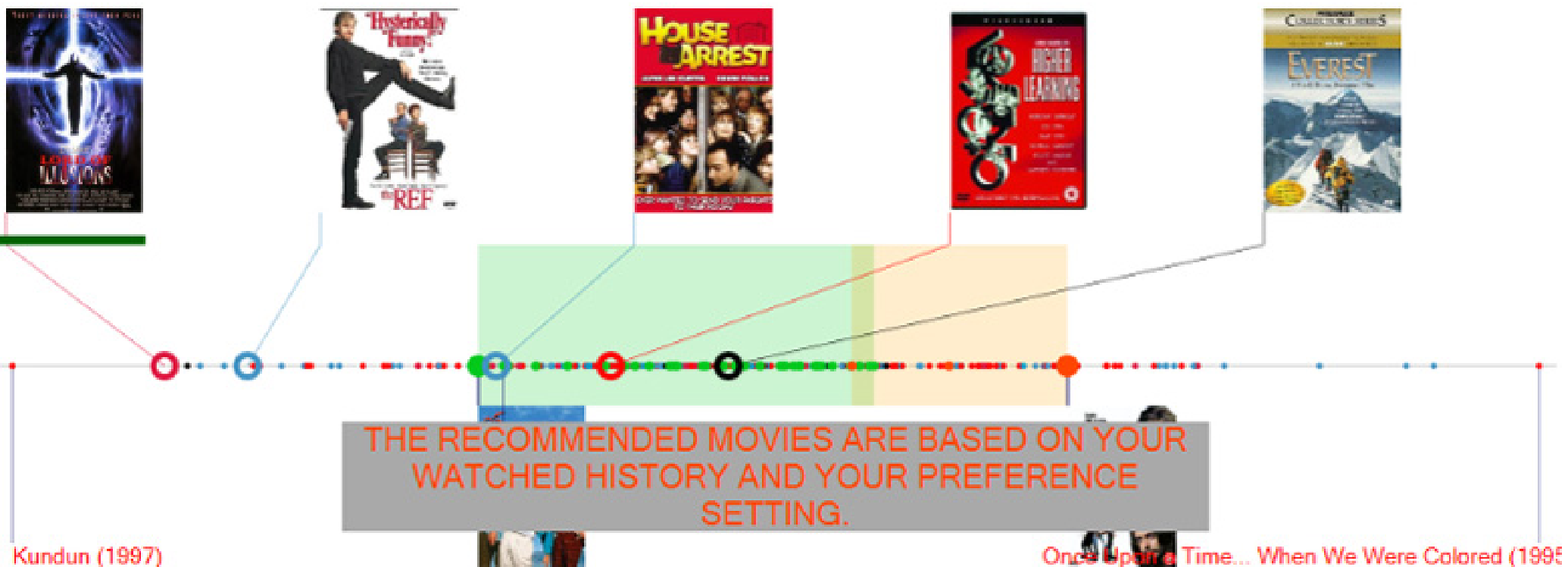} \ \ \
\includegraphics[width = 1.64in]{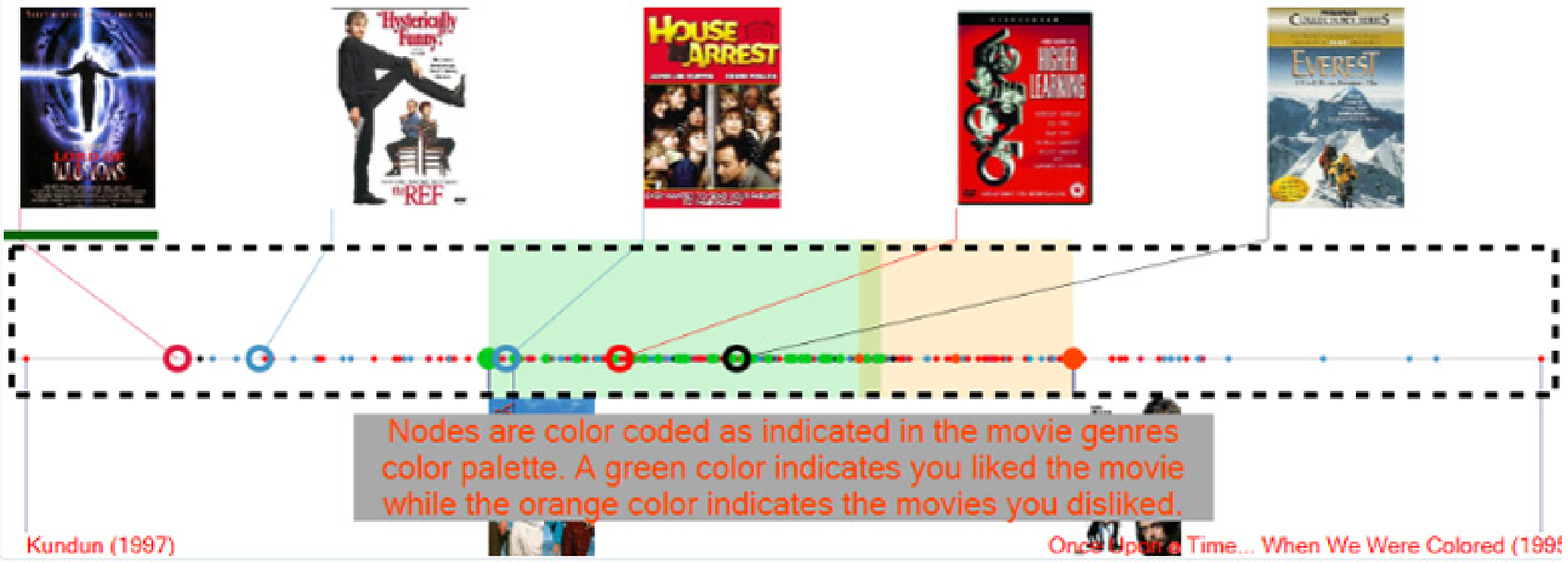} \ \ \
\includegraphics[width = 1.64in]{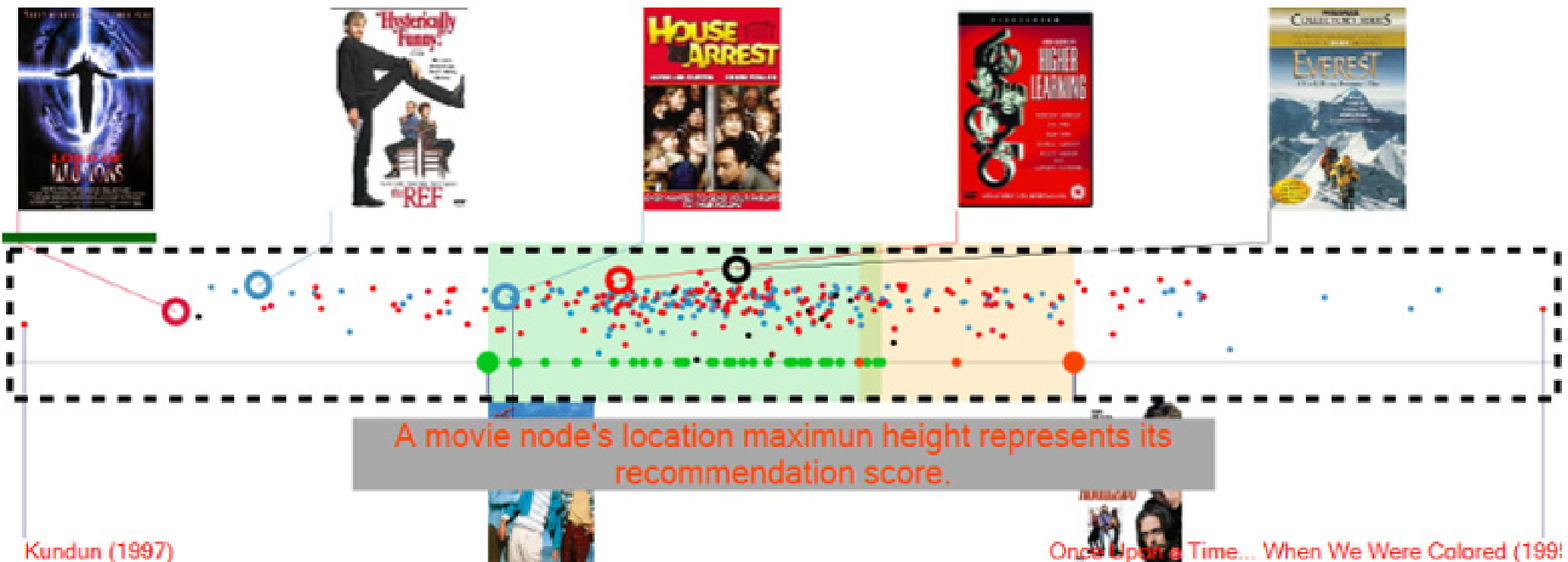} \ \ \
\includegraphics[width = 1.64in]{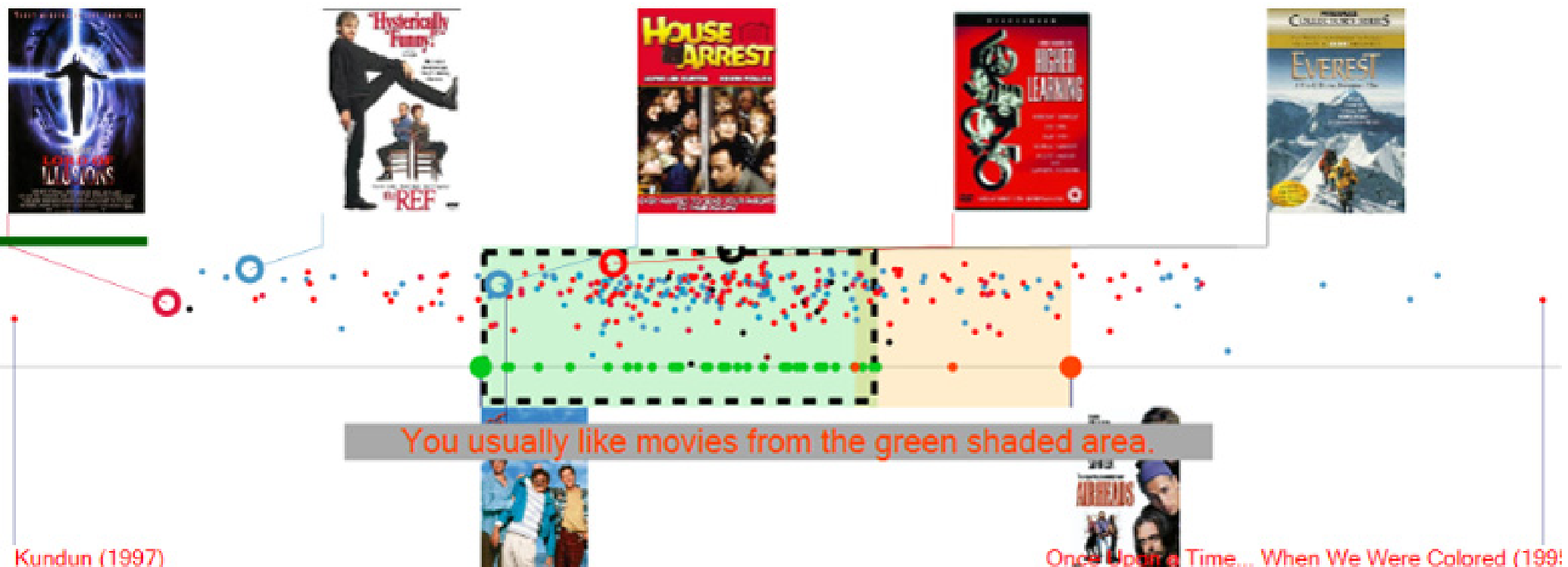}\\ 
 \ \ \ \ \ \ (a) \ \ \ \ \ \ \ \ \ \ \ \ \ \ \ \ \ \ \ \ \ \ \ \ \ \ \ \ \ \ \ \ \ \ \ \ \ \ \ \  \ \ \ \ \ \ \ \ \ \ \ \ (b) \ \ \ \ \ \ \ \ \ \ \ \ \ \ \ \ \ \ \ \ \ \ \ \ \ \ \ \ \ \ \ \ \ \ \ \ \ \ \ \  \ \ \ \ \ \ \ \ \ \ \ \ (c) \ \ \ \ \ \ \ \ \ \ \ \ \ \ \ \ \ \ \ \ \ \ \ \ \ \ \ \ \ \ \ \ \ \ \ \ \ \ \ \  \ \ \ \ \ \ \ \ \ \ \ \ (d) \\ 
\includegraphics[width = 1.64in]{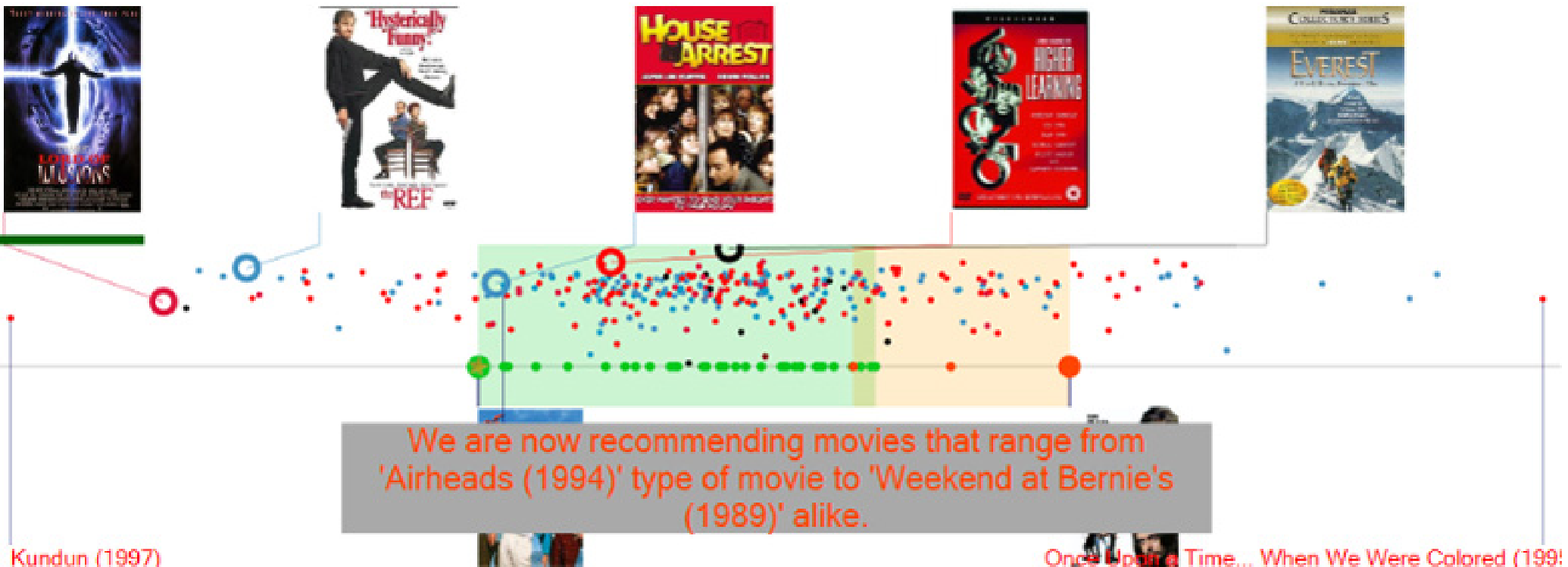} \ \ \
\includegraphics[width = 1.64in]{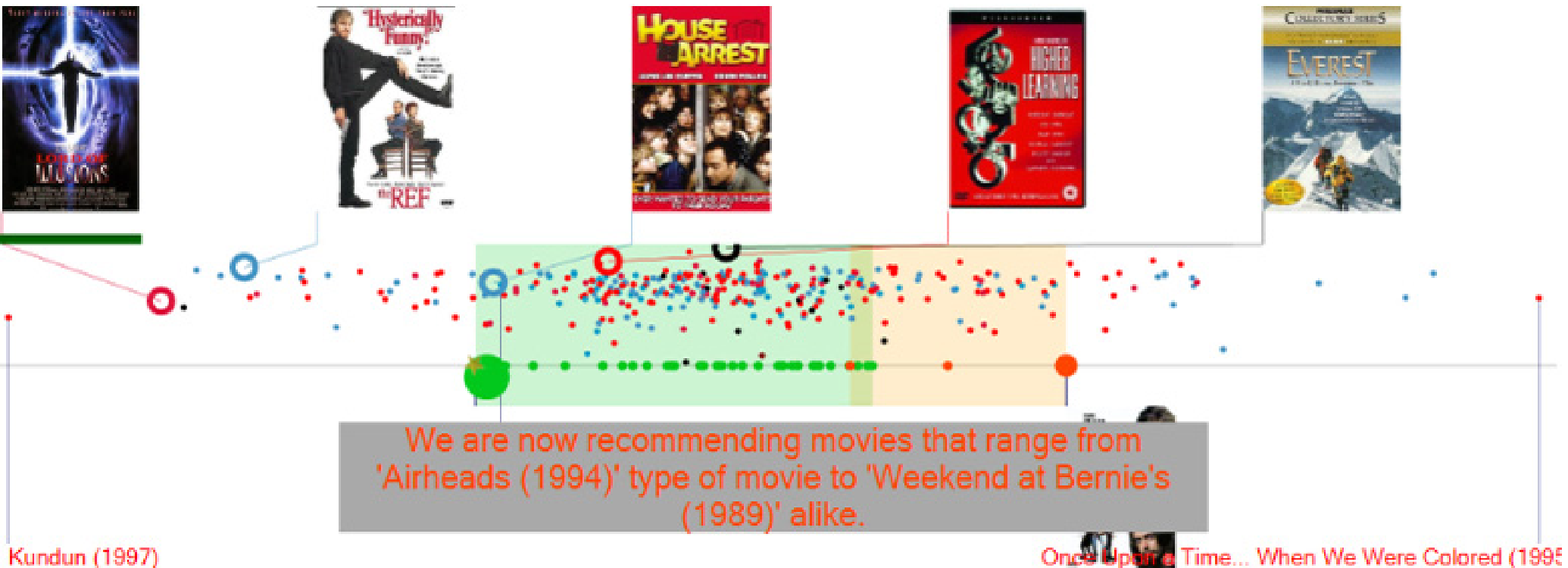} \ \ \
\includegraphics[width = 1.64in]{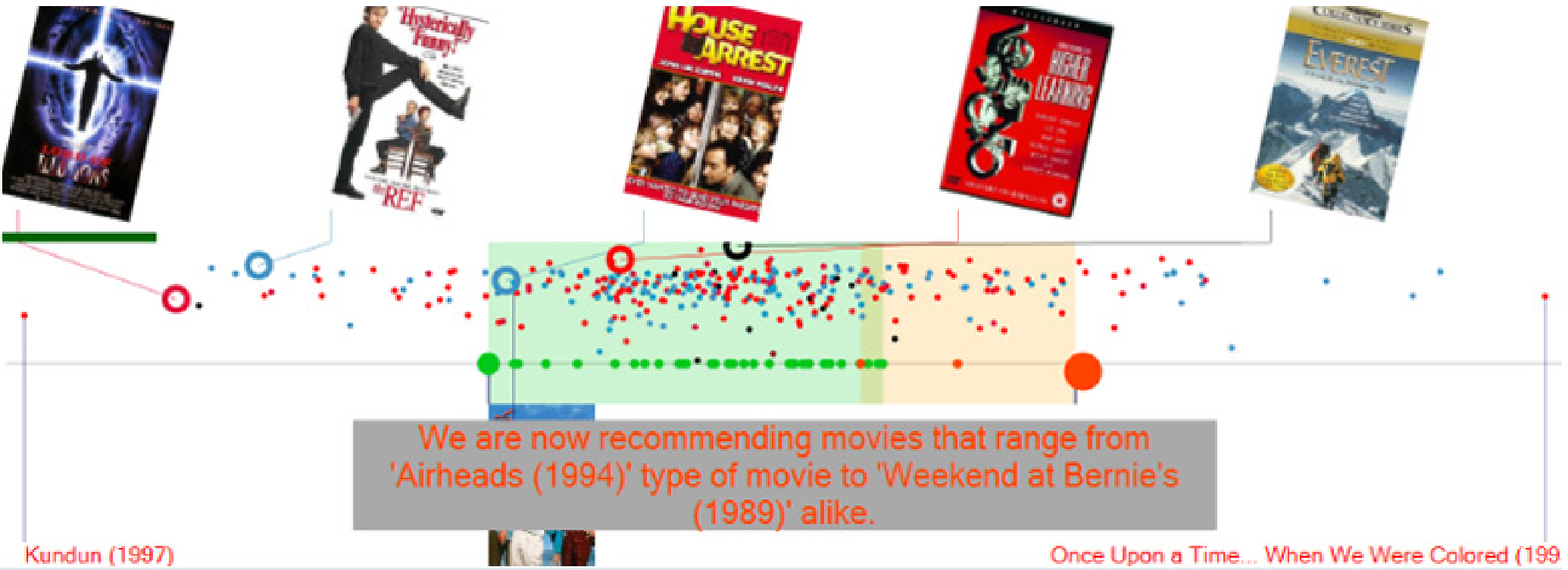} \ \ \
\includegraphics[width = 1.64in]{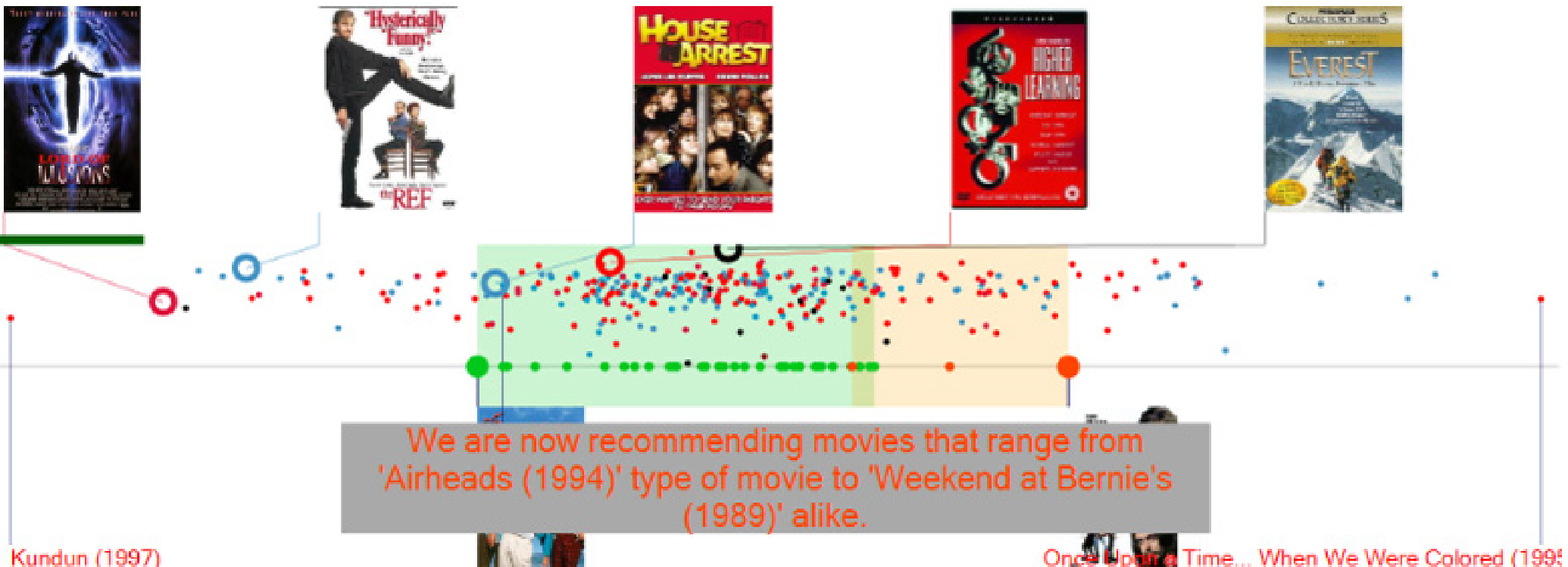}\\
  \ \ \ \ \ \ (e) \ \ \ \ \ \ \ \ \ \ \ \ \ \ \ \ \ \ \ \ \ \ \ \ \ \ \ \ \ \ \ \ \ \ \ \ \ \ \ \  \ \ \ \ \ \ \ \ \ \ \ \ (f) \ \ \ \ \ \ \ \ \ \ \ \ \ \ \ \ \ \ \ \ \ \ \ \ \ \ \ \ \ \ \ \ \ \ \ \ \ \ \ \  \ \ \ \ \ \ \ \ \ \ \ \ (g) \ \ \ \ \ \ \ \ \ \ \ \ \ \ \ \ \ \ \ \ \ \ \ \ \ \ \ \ \ \ \ \ \ \ \ \ \ \ \ \  \ \ \ \ \ \ \ \ \ \ \ \ (h) \\ 
\includegraphics[width = 1.64in]{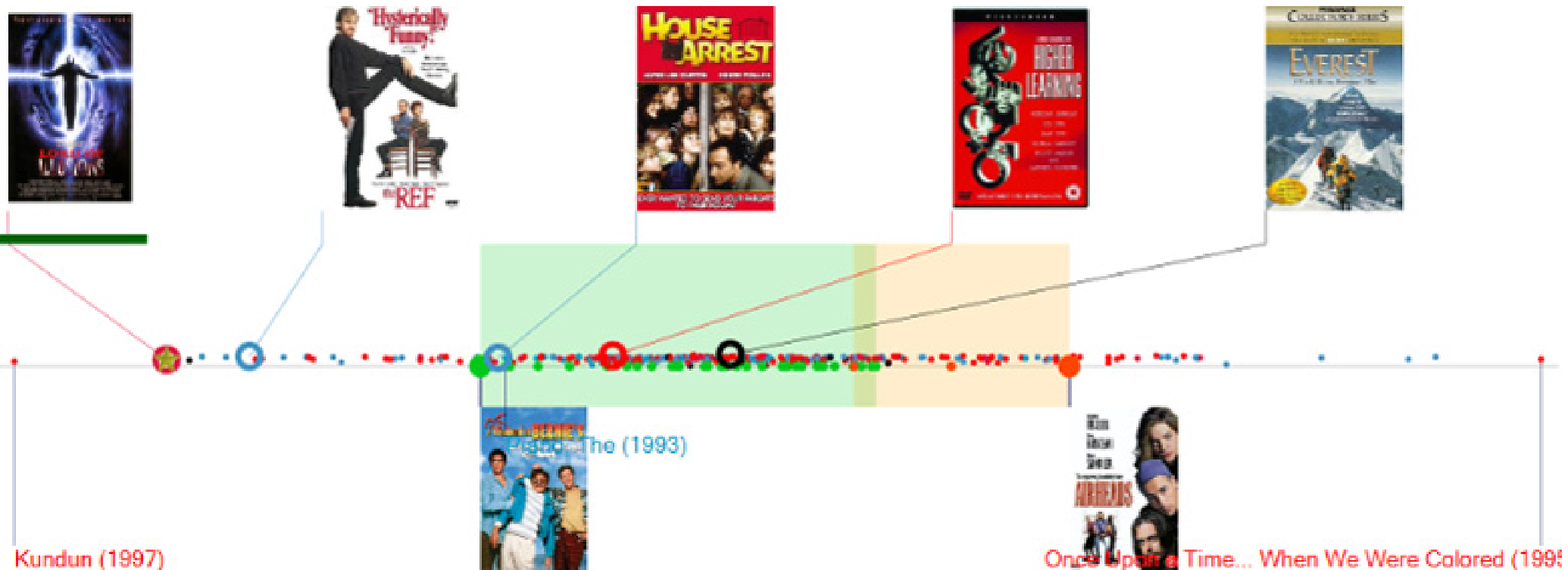} \ \ \
\includegraphics[width = 1.64in]{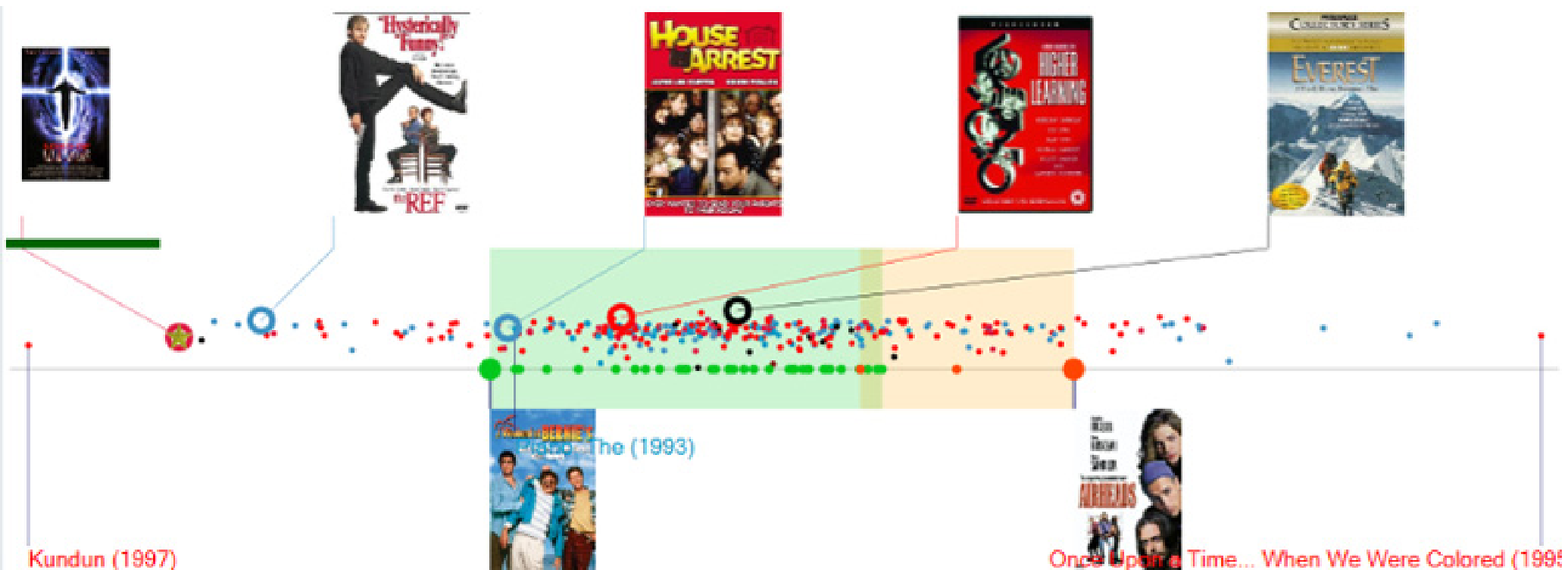} \ \ \
\includegraphics[width = 1.64in]{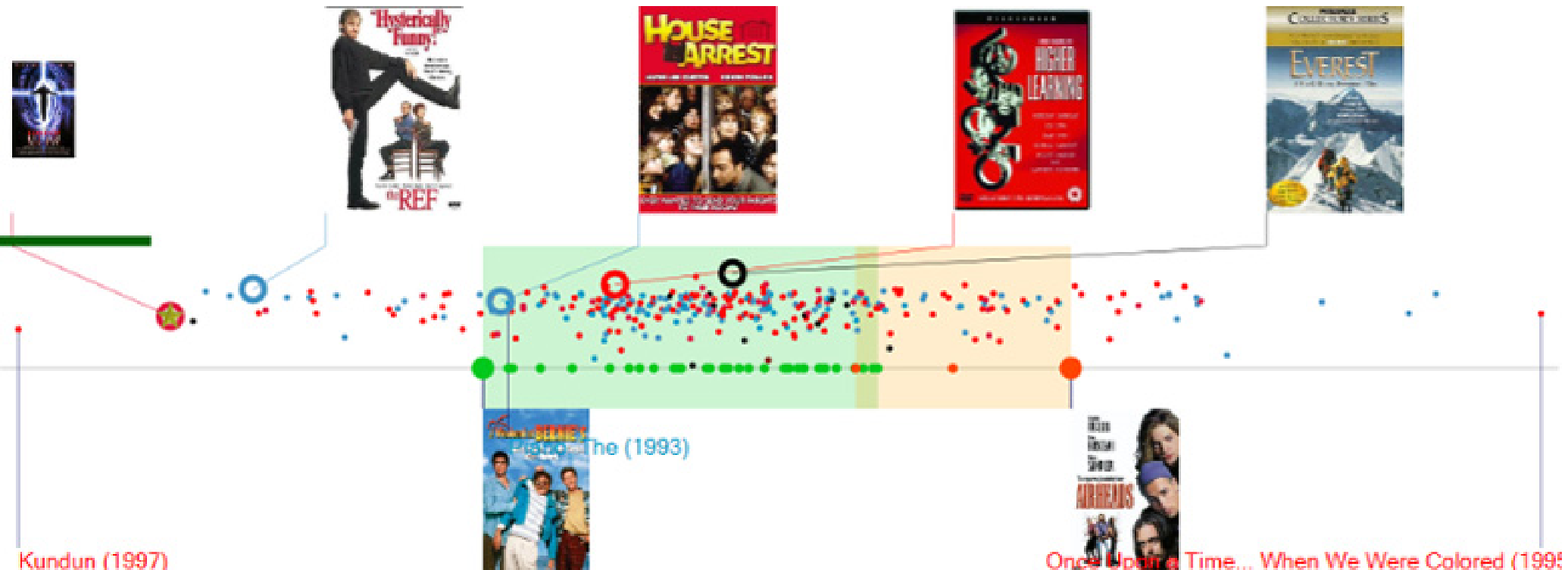} \ \ \
\includegraphics[width = 1.64in]{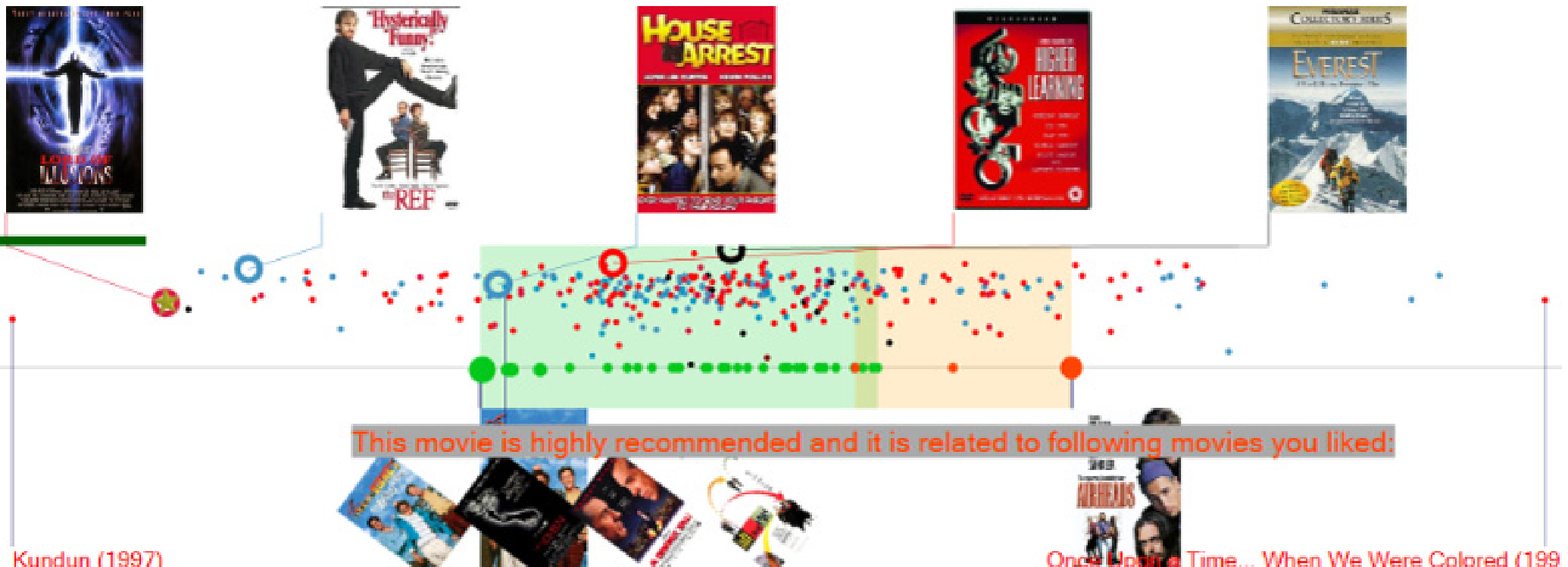}\\
 \ \ \ \ \ \ (i)  \ \ \ \ \ \ \ \ \ \ \ \ \ \ \ \ \ \ \ \ \ \ \ \ \ \ \ \ \ \ \ \ \ \ \ \ \ \ \ \  \ \ \ \ \ \ \ \ \ \ \ \ (j) \ \ \ \ \ \ \ \ \ \ \ \ \ \ \ \ \ \ \ \ \ \ \ \ \ \ \ \ \ \ \ \ \ \ \ \ \ \ \ \  \ \ \ \ \ \ \ \ \ \ \ \ (k) \ \ \ \ \ \ \ \ \ \ \ \ \ \ \ \ \ \ \ \ \ \ \ \ \ \ \ \ \ \ \ \ \ \ \ \ \ \ \ \  \ \ \ \ \ \ \ \ \ \ \ \ (l) \\ 
\includegraphics[width = 1.64in]{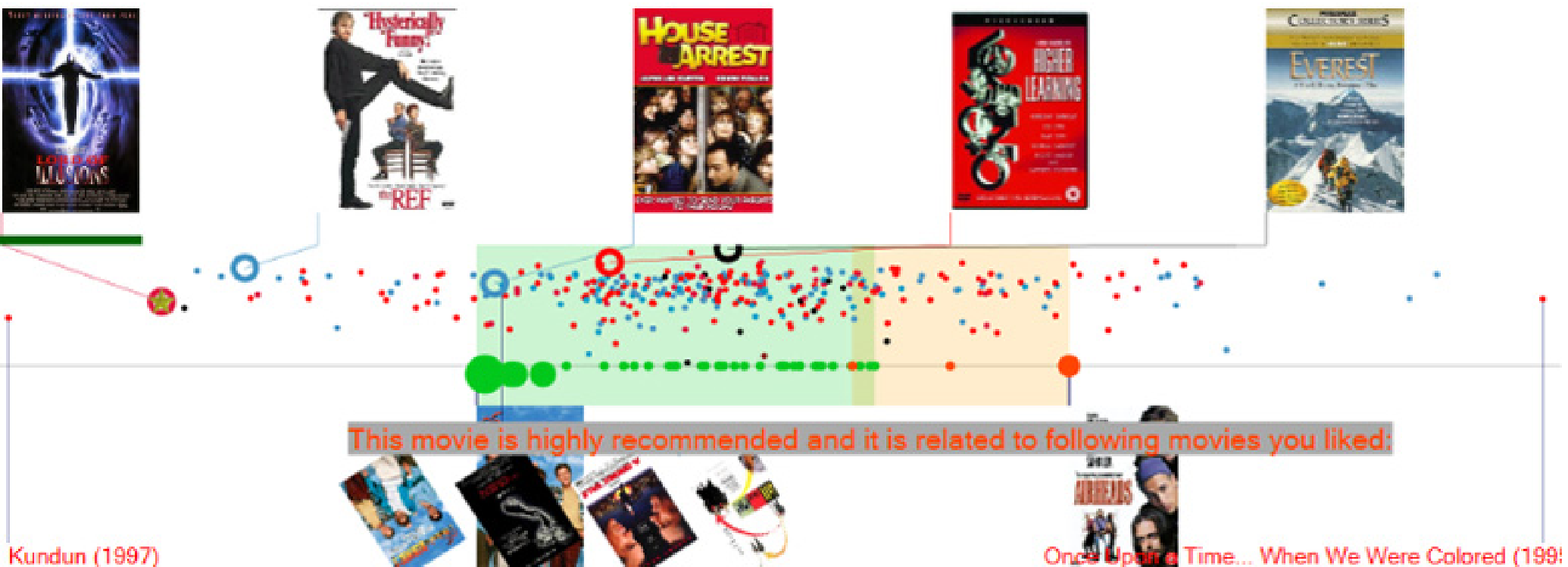} \ \ \
\includegraphics[width = 1.64in]{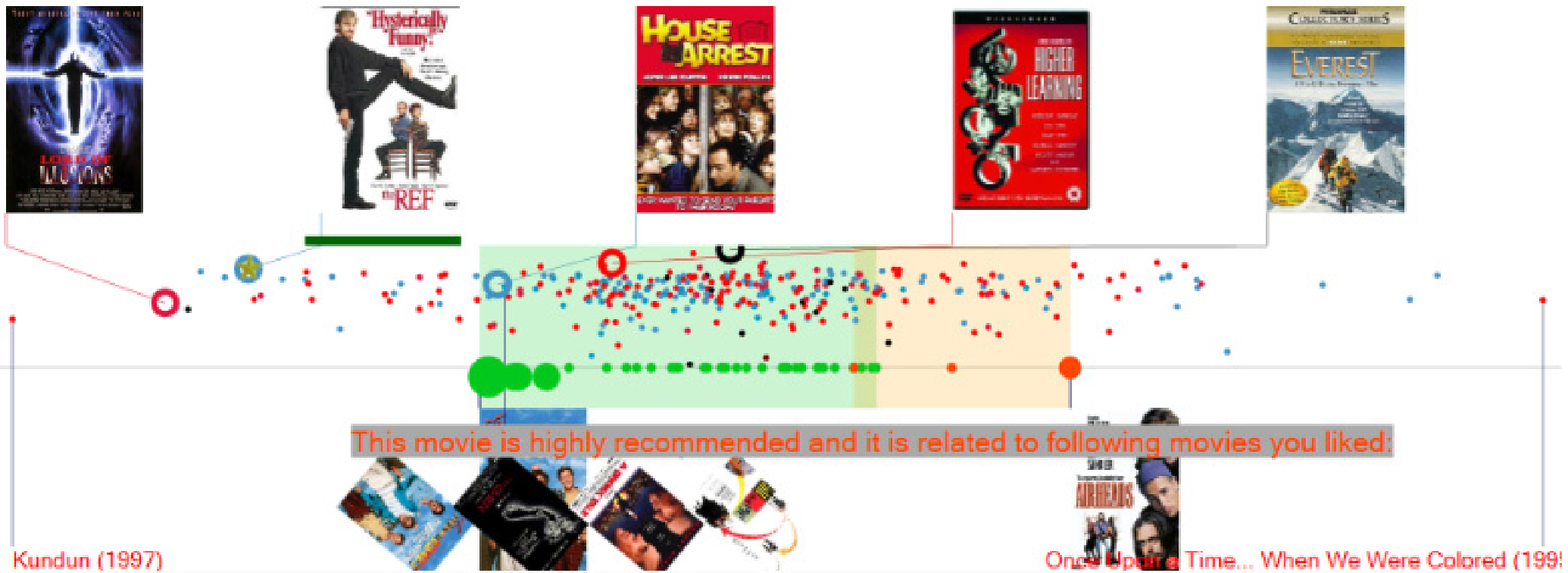} \ \ \
\includegraphics[width = 1.64in]{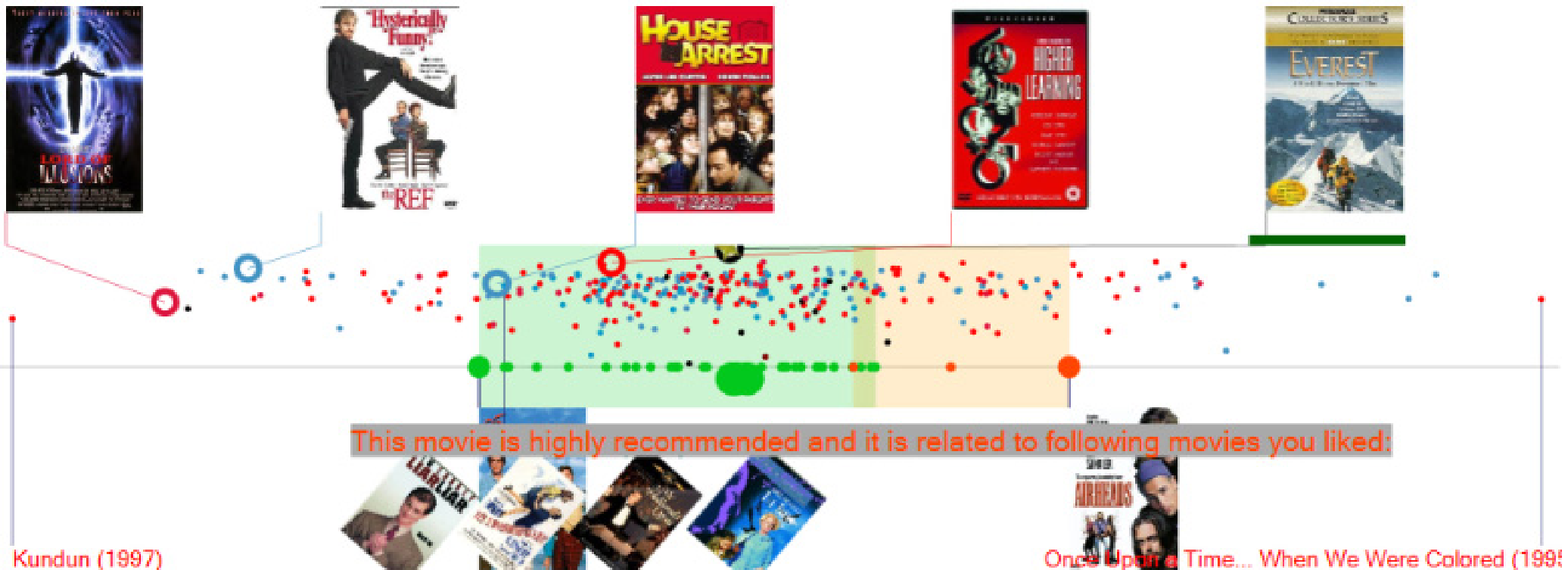}\\
  \ \ \ \ \ \ (m) \ \ \ \ \ \ \ \ \ \ \ \ \ \ \ \ \ \ \ \ \ \ \ \ \ \ \ \ \ \ \ \ \ \ \ \ \ \ \ \  \ \ \ \ \ \ \ \ \ \ \ \ (n) \ \ \ \ \ \ \ \ \ \ \ \ \ \ \ \ \ \ \ \ \ \ \ \ \ \ \ \ \ \ \ \ \ \ \ \ \ \ \ \  \ \ \ \ \ \ \ \ \ \ \ \ (o)\\ 
\caption{Example snapshots of the storytelling animation. 
We start by introducing the system interface to a new user, such as the use of rating history (a), color code (b), recommendation degree (c) and the liked zone (d).
We also use example movies to illustrate the latent dimension and attract user attention with poster transitions (e-h).
Next, we introduce the first recommended movie by animating the movie node and showing a green line under its poster (i).
The visualization domain is animated from level one (i), level two (j-k), to level three (l-m) gradually.
The same procedure (i-m) is used to animate the second (n) to the last (o) recommended movies respectively.
}


\label{animation}
\end{figure*}

As shown in image (i) of Figure~\ref{animation}, the level one provides the basic information of a latent dimension with the liked and disliked regions, where the nodes of movies being recommended are drawn as circles.
The nodes of movies that the user has watched are colored green and the nodes of recommendable movies are colored based on their genres.
Two example movies are provided for illustrating the latent dimension and making correlations to other movies.


The level two introduces the recommendation degrees to the vertical locations of recommendable movies. As shown in images (j) and (k) of Figure~\ref{animation}, the movies with higher recommendation degrees are placed on the top to attract user attention. 
The scaling can be automatically done in an animation or interactively adjusted by the user when exploring the movie database.

The level three provides the richest information for exploring movies.
For the movie being recommended, we reveal the top four similar movies that the user has watched and liked to strengthen the reasons of recommendation, as shown in images (l) and (m)  of Figure~\ref{animation}.
All the four movies have high user ratings and they are close to the movie being recommended on the latent dimension.
We also color all the movie nodes based on the genres and add colored links to connect posters to the movie nodes.


\subsection{Animation Effects}

Our system provides fully automatic animations to ``tell" a recommendation story with the following three sets of animation effects.

The first animation set is to introduce the system components. 
Step by step, each component is highlighted with a brief description as shown in images (a-d) in Figure~\ref{animation}. 
The user can replay or skip to the next animation set anytime.
 
The second animation set is to present the narrative structure of a recommendation story.
We use our example-based approach to animate the recommended movies for user attention and the two example movies from left to right for providing the impression of dimension meanings.
Specifically, each example movie poster is animated by changing its size and the movie node is also animated simultaneously, as shown in  Figure~\ref{animation} (e-h).


The third animation set is to present a recommendation story, as shown in images (i-o) in Figure~\ref{animation}.
We start to highlight the selected movie by flashing the movie node in the visualization panel when its poster in the recommendation list is animated.  
A green line under the movie poster also indicates the focused movie. 
Then, we switch the visualization from level one to level three gradually to provide detailed information of a selected movie.
Specifically, after the level one is shown, the movie nodes in the visualization panel are progressively scaled to their maximum recommendation degree in the level two.
From level two to level three, the set of similar movies are animated.
We use the same animation procedure for all the selected movies in the story to avoid confusion.
The user can observe all the information or switch anytime to the next recommended movie.

When the animation of a recommendation story ends or the user selects a movie for further exploration, a new story is generated automatically.
The system re-introduces the new narrative structure and repeats the animation sequences to convey the story behind the new selected movies.





\section{Results and Case Studies}

\subsection{Case Studies}

This section describes three case studies using the MovieLens100K dataset~\cite{MovieLens100k}. 
The dataset has 100K ratings from 1 to 5 and 1682 movies from different categories rated by 943 users.
We have select users with different backgrounds and rating histories to demonstrate our approach.






\textbf{The first user} is a 21 year old male working in entertainment.
As shown in Figure~\ref{interface}, a latent dimension is identified between comedy movies toward the right and drama/romance/biography genres toward the left.
As the default preference settings of 0.5/0.5 for typicality/un-typicality and familiarity/diversity, our recommendation system starts with movies of drama genre in the familiar zone and continues to drama and romance types for diversity. 
The recommended movies are: When a Man Loves a Woman (1994, Drama and Romance), Meet John Doe (1941, Drama, Romance and Comedy), Nobody’s Fool (1994, Drama and Comedy), Ref The (1994, Comedy and Drama), and Rebel Without a Cause (1995, Drama).

We compare the recommendation results with the user's watch history.
The user has rated 44 movies.
Among which, about 68\% of the movies are drama or a combination of drama and romance genres. Their average rating is high (4.30).
The movies of other genres received lower ratings, especially comedy movies.
Therefore, our result captures the user's preference of drama movies over comedy types.

\begin{figure}[htb]
\centering
\includegraphics[width = 3.3in]{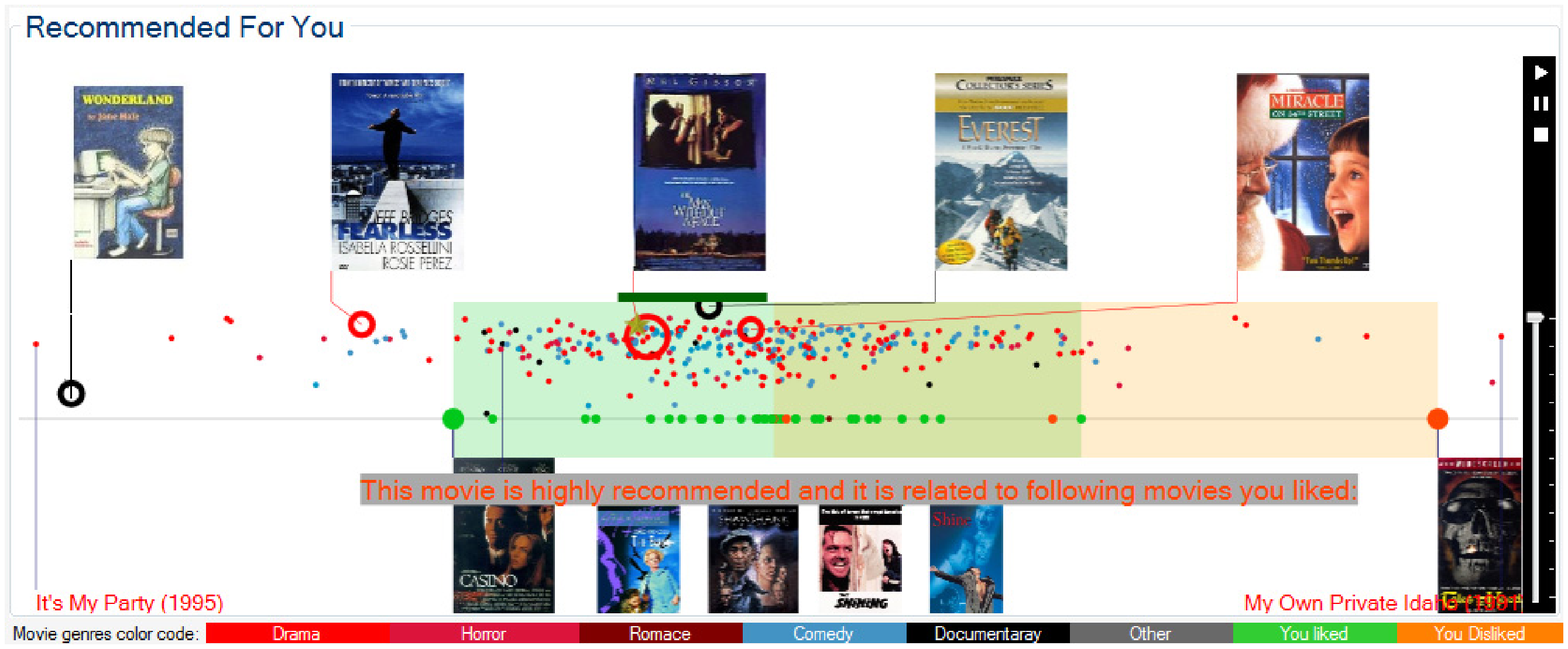}
\caption{The result of the second user describes the user preference of love/drama movies he liked toward the left to horror/comedy movies he disliked toward the right.}
\label{case1RecommendedList}
\end{figure}

\textbf{The second user} is a 28 year old male student.
Figure~\ref{case1RecommendedList} shows the first recommendation results for the user. 
The latent dimension spreads out movies in drama, documentary and comedy genres, with the combined features of positive influence (such as love and drama) toward the left direction and negative influence (such as horror) toward the right. 
Five movies are recommended from diverse to familiar zones: Wonderland (1997, Documentary), Fearless (1993, Drama), Man Without a Face (The 1993, Drama), Everest (1998, Documentary), and Miracle on 34th Street (1994, Drama).

\begin{figure}[htb]
\centering
\includegraphics[width = 3.3in]{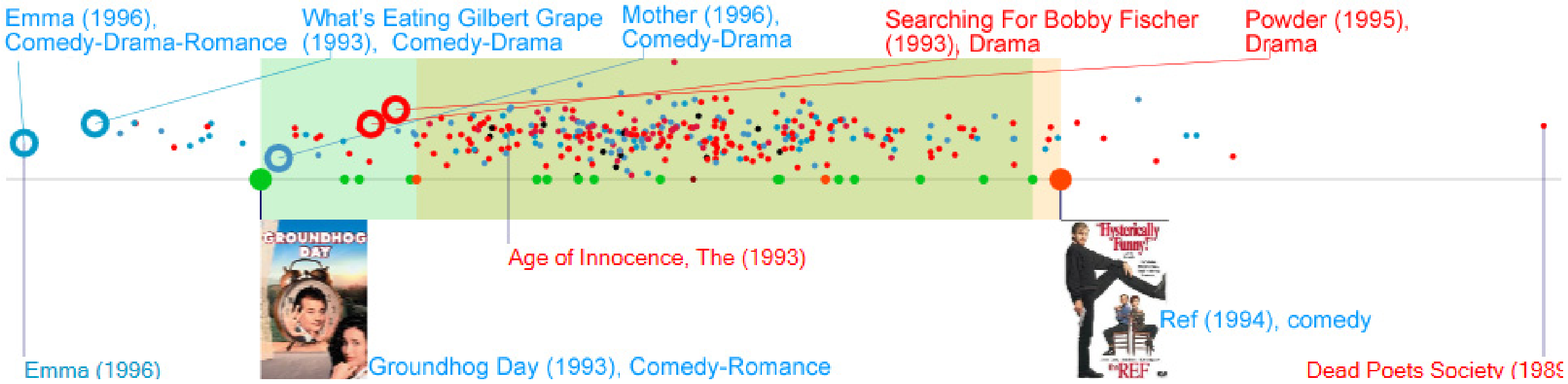}\\
\vspace{+1mm}
\includegraphics[width = 3.3in]{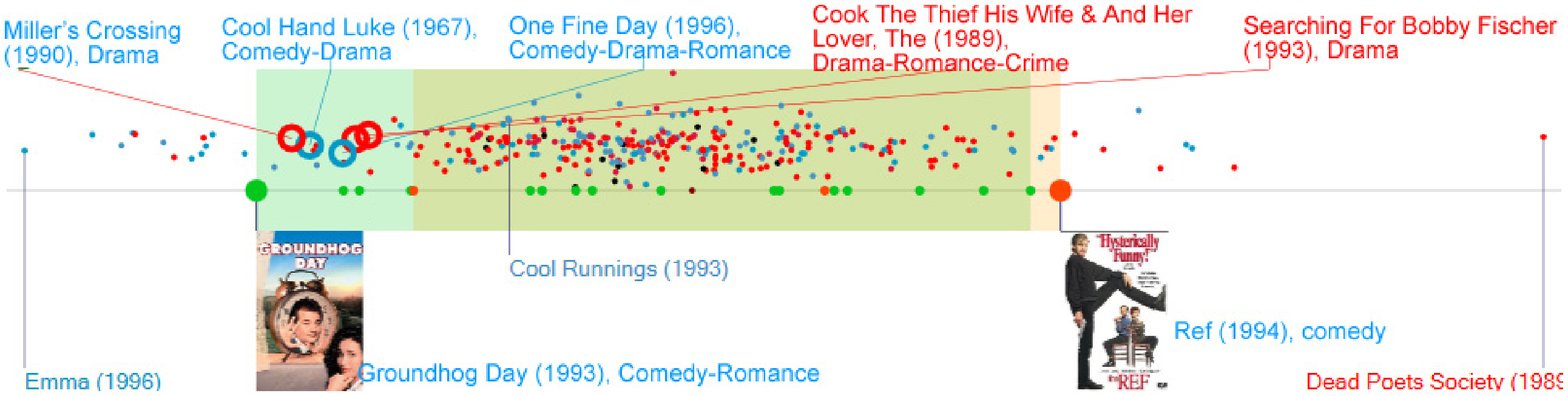}\\
\vspace{+1mm}
\includegraphics[width = 3.3in]{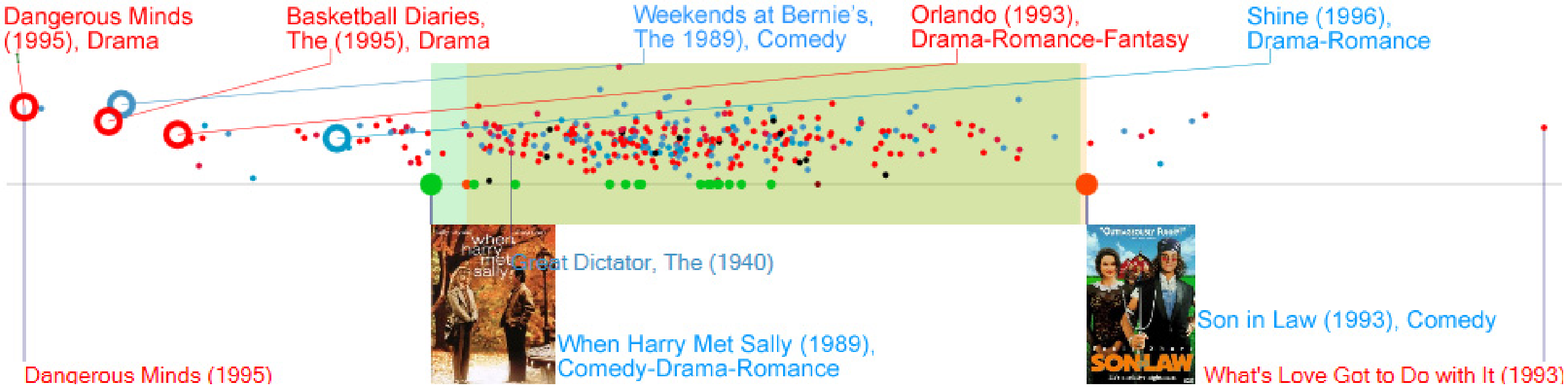}
\caption{Interaction examples for the second user.
The first row shows another story piece describing a different aspect of user's preference on Groundhog Day (1993, Comedy and Romance) over Son in Law (1993, Comedy).
The second row shows that all recommended movies come from the familiar zone when the user switch the familiar preference to 1.
The bottom row shows the result when both the typical and diverse preferences are set to 1 and the movie ``One Fine Day (1996)'' from the recommended list is selected.
}
\label{case1InteractionsResults}
\end{figure}

Without user input, our system continues to an additional story with a new set of movies as shown in the first row of Figure~\ref{case1InteractionsResults}.
The user can also adjust the recommendation preferences and select their interested movies, shown on the second and third rows of Figure~\ref{case1InteractionsResults}.  

We compare the recommendation results with the user's watch history.
He has an average rating of 3.64 for 34 movies from comedy, drama, horror, and romance genres.
The rating records suggest that he likes drama and romance movies. 
Our recommendation results recommend majority movies in the familiar zone of the user and also introduce mixed adventure/documentary/drama movies for diversity.

\begin{figure}[htb]
\centering
\includegraphics[width = 3.3in]{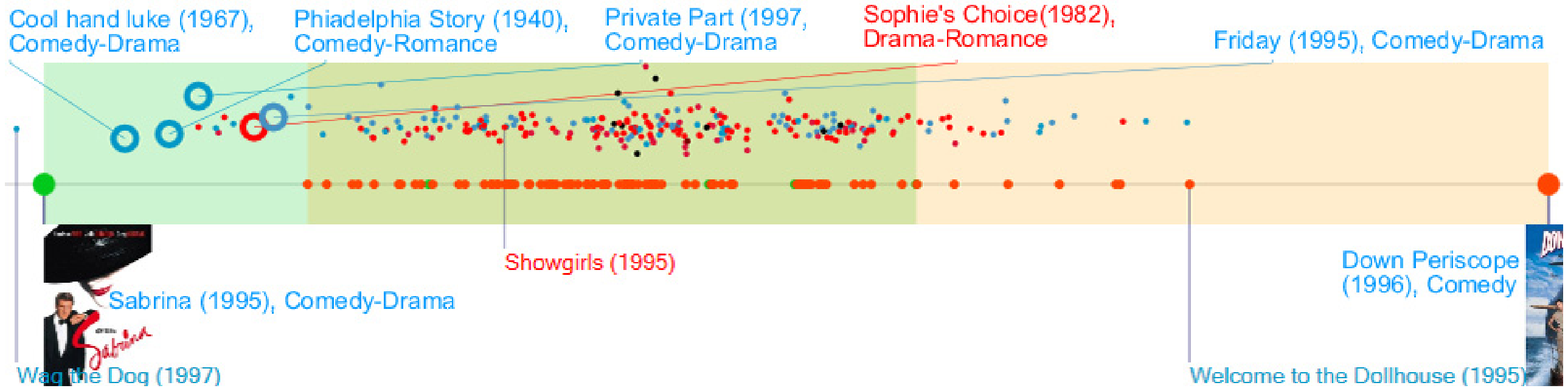}
\caption{The result of the third user demonstrates a picky user who has rated many movies low.
Our approach identifies the user's preference on comedy-drama movies (toward the left) over the other genres (toward the right).
}
\label{case2RecommendedList}
\end{figure}

\textbf{The third user} is a 26 year old male executive.
As shown in Figure~\ref{case2RecommendedList}, the latent dimension is delimited by drama-comedy movie ``Sabrina (1996)'' with a high rating on the left and the comedy movie ``Down Periscope (1996)'' with a low rating on the right. 
Our system recommends five movies from drama movies combined with other genres in the familiar zone, such as Cool Hand Luke (1967, Comedy and Drama), Philadelphia Story (1940, comedy romance), Private Part (1997, Comedy and Drama), Sophie’s Choice (1982, Drama and Romance) and Friday (1995, Comedy and Drama).

We compare the recommendation results with the user's watch history.
He has rated 120 movies of several genres, such as comedy, drama, and horror.
Most of his ratings were 1 or 2 stars and he only rated six movies with 4 or 5 stars.
Our result reflects his low rating records with a large dislike region and a small like region.
The latent dimension describes movie types from drama-comedy movies toward the left, to various other movie genres that the user has rated low toward the right.
Such users are generally picky on movie selections, but our model still captures his favorite movie types and creates a matching recommendation story.

\subsection{System Performance}

The preprocess of our system includes the computation described in Section 3, including process of rating records, construction of SVD space, selection of recommendable movies and latent dimensions for the user. 
The performance of this stage is depended on the sizes of movie database and rating records.
The preprocess takes 3-10 seconds for 30,000 ratings from 940 users for 370 movies and several minutes for the Movie100K dataset on a desktop computer with Intel Core i7 2.93 GHz processor.

During interactive recommendation stage, all the visualization, interaction, and storytelling processes described in Section 4 and 5 are interactive. 
This is essential for providing smooth user interaction in an online system.
It is achieved by only keeping the relevant data to the user during run time.
The performance is the main reason that rating records are used in our approach and many other popular online systems.



\section{Conclusions and Future Work}

This paper presents an interactive recommendation approach for the popular application of online movie recommendation with the general public as end users.
We have studied LSM, which enables us to translate abstract data and complex recommendation algorithms to a set of semantic concepts, provide explicit interaction of search preferences, and construct meaningful recommendation stories.
The LSM can be easily combined with other recommendation algorithms, as we separate the estimation of recommendation degrees and the latent space.
The interactive recommendation approach automatically generates storytelling animations for recommendation, with the supports of several interactive exploration functions for users to adjust the search results explicitly. 
Different from traditional recommendation algorithms, the interactive recommendation approach emphasizes the visual communication between users and recommendation systems for engaging users and improving search experiences.
Both of our results can also be extended to recommend other online products or services.

As future work, we plan to perform formal evaluations on the effectiveness of interactive recommendation.
We are interested in studying the suitable amount of information for different users, such as the general public and movie experts, so that different versions of narrative visualization can be developed to suit for different needs. 
We also plan to develop variations of recommendation stories, such as long versions that can combine multiple latent dimensions, for different recommendation tasks.
At the end, we are interested in integrating other techniques for movie recommendation, such as text analysis approaches for mining useful information from the movie reviews.
The results will enrich the contents of narrative visualization and may provide better search experiences.




\bibliographystyle{abbrv}
\bibliography{story}

\begin{thebibliography}{10}

\bibitem{amini2015understanding}
F.~Amini, N.~Henry~Riche, B.~Lee, C.~Hurter, and P.~Irani.
\newblock Understanding data videos: Looking at narrative visualization through
  the cinematography lens.
\newblock In {\em Proceedings of the 33rd Annual ACM Conference on Human
  Factors in Computing Systems}, pages 1459--1468. ACM, 2015.

\bibitem{bach2016telling}
B.~Bach, N.~Kerracher, K.~W. Hall, S.~Carpendale, J.~Kennedy, and N.~H. Riche.
\newblock Telling stories about dynamic networks with graph comics.
\newblock In {\em Proceedings of the Conference on Human Factors in Information
  Systems (CHI)}. ACM, New York, United States, 2016.

\bibitem{boy2015storytelling}
J.~Boy, F.~Detienne, and J.-D. Fekete.
\newblock Storytelling in information visualizations: Does it engage users to
  explore data?
\newblock In {\em Proceedings of the ACM Conference on Human Factors in
  Computing Systems}, CHI '15, pages 1449--1458, 2015.

\bibitem{7539294}
C.~Bryan, K.~L. Ma, and J.~Woodring.
\newblock Temporal summary images: An approach to narrative visualization via
  interactive annotation generation and placement.
\newblock {\em IEEE Transactions on Visualization and Computer Graphics}, 2016.

\bibitem{crnovrsanin2011visual}
T.~Crnovrsanin, I.~Liao, Y.~Wuy, and K.-L. Ma.
\newblock Visual recommendations for network navigation.
\newblock In {\em Proceedings of the 13th Eurographics / IEEE - VGTC Conference
  on Visualization}, EuroVis'11, pages 1081--1090, 2011.

\bibitem{cruz2011generative}
P.~Cruz and P.~Machado.
\newblock Generative storytelling for information visualization.
\newblock {\em IEEE computer graphics and applications}, 31(2):80--85, 2011.

\bibitem{Deshpande:2004:ITN:963770.963776}
M.~Deshpande and G.~Karypis.
\newblock Item-based top-n recommendation algorithms.
\newblock {\em ACM Trans. Inf. Syst.}, 22(1):143--177, Jan. 2004.

\bibitem{dragicevic2011temporal}
P.~Dragicevic, A.~Bezerianos, W.~Javed, N.~Elmqvist, and J.-D. Fekete.
\newblock Temporal distortion for animated transitions.
\newblock In {\em Proceedings of the SIGCHI Conference on Human Factors in
  Computing Systems}, pages 2009--2018. ACM, 2011.

\bibitem{4388992}
R.~Eccles, T.~Kapler, R.~Harper, and W.~Wright.
\newblock Stories in geotime.
\newblock In {\em Visual Analytics Science and Technology, 2007. VAST 2007.
  IEEE Symposium on}, pages 19--26, 2007.

\bibitem{Ekstrand:2011:CFR:2185827.2185828}
M.~D. Ekstrand, J.~T. Riedl, and J.~A. Konstan.
\newblock Collaborative filtering recommender systems.
\newblock {\em Found. Trends Hum.-Comput. Interact.}, 4(2):81--173, Feb. 2011.

\bibitem{6902874}
A.~Figueiras.
\newblock How to tell stories using visualization.
\newblock In {\em Information Visualisation (IV), 2014 18th International
  Conference on}, pages 18--26, 2014.

\bibitem{4677364}
D.~Fisher, A.~Hoff, G.~Robertson, and M.~Hurst.
\newblock Narratives: A visualization to track narrative events as they
  develop.
\newblock In {\em IEEE Symposium on Visual Analytics Science and Technology},
  pages 115--122, 2008.

\bibitem{Gershon:2001:SIV:381641.381653}
N.~Gershon and W.~Page.
\newblock What storytelling can do for information visualization.
\newblock {\em Commun. ACM}, 44(8):31--37, Aug. 2001.

\bibitem{Gomez-Uribe:2015:NRS:2869770.2843948}
C.~A. Gomez-Uribe and N.~Hunt.
\newblock The netflix recommender system: Algorithms, business value, and
  innovation.
\newblock {\em ACM Trans. Manage. Inf. Syst.}, 6(4):13:1--13:19, Dec. 2015.

\bibitem{gretarsson2010smallworlds}
B.~Gretarsson, J.~O'Donovan, S.~Bostandjiev, C.~Hall, and T.~H{\"o}llerer.
\newblock Smallworlds: Visualizing social recommendations.
\newblock {\em Computer Graphics Forum}, 29(3):833--842, 2010.

\bibitem{6231611}
X.~Hu, A.~Lu, and X.~Wu.
\newblock Spectrum-based network visualization for topology analysis.
\newblock {\em Computer Graphics and Applications, IEEE}, 33(1):58--68, Jan
  2013.

\bibitem{hullman2011visualization}
J.~Hullman and N.~Diakopoulos.
\newblock Visualization rhetoric: Framing effects in narrative visualization.
\newblock {\em Visualization and Computer Graphics, IEEE Transactions on},
  17(12):2231--2240, 2011.

\bibitem{Hullman:2013:CAG:2470654.2481374}
J.~Hullman, N.~Diakopoulos, and E.~Adar.
\newblock Contextifier: Automatic generation of annotated stock visualizations.
\newblock In {\em Proceedings of the SIGCHI Conference on Human Factors in
  Computing Systems}, CHI '13, pages 2707--2716, 2013.

\bibitem{Hullman:2013:DUS:2553699.2553753}
J.~Hullman, S.~Drucker, N.~Henry~Riche, B.~Lee, D.~Fisher, and E.~Adar.
\newblock A deeper understanding of sequence in narrative visualization.
\newblock {\em IEEE Transactions on Visualization and Computer Graphics},
  19(12):2406--2415, 2013.

\bibitem{kermarrec2012data}
A.-M. Kermarrec and A.~Moin.
\newblock {Data Visualization Via Collaborative Filtering}.
\newblock Research report, Inria, Feb. 2012.

\bibitem{Koren:2008:FMN:1401890.1401944}
Y.~Koren.
\newblock Factorization meets the neighborhood: A multifaceted collaborative
  filtering model.
\newblock In {\em Proceedings of the 14th ACM SIGKDD International Conference
  on Knowledge Discovery and Data Mining}, KDD '08, pages 426--434, 2008.

\bibitem{6412677}
R.~Kosara and J.~Mackinlay.
\newblock Storytelling: The next step for visualization.
\newblock {\em Computer}, 46(5):44--50, 2013.

\bibitem{Lee:2013:STM:2553699.2553755}
B.~Lee, R.~H. Kazi, and G.~Smith.
\newblock Sketchstory: Telling more engaging stories with data through freeform
  sketching.
\newblock {\em IEEE Transactions on Visualization and Computer Graphics},
  19(12):2416--2425, Dec. 2013.

\bibitem{7274435}
B.~Lee, N.~Riche, P.~Isenberg, and S.~Carpendale.
\newblock More than telling a story: Transforming data into visually shared
  stories.
\newblock {\em Computer Graphics and Applications, IEEE}, 35(5):84--90, 2015.

\bibitem{leskovec2014mining}
J.~Leskovec, A.~Rajaraman, and J.~D. Ullman.
\newblock {\em Mining of massive datasets}.
\newblock Cambridge University Press, 2014.

\bibitem{1167344}
G.~Linden, B.~Smith, and J.~York.
\newblock Amazon.com recommendations: item-to-item collaborative filtering.
\newblock {\em IEEE Internet Computing}, 7(1):76--80, 2003.

\bibitem{loepp2015blended}
B.~Loepp, K.~Herrmanny, and J.~Ziegler.
\newblock Blended recommending: Integrating interactive information filtering
  and algorithmic recommender techniques.
\newblock In {\em Proceedings of ACM Conference on Human Factors in Computing
  Systems}, pages 975--984. ACM, 2015.

\bibitem{loepp2014choice}
B.~Loepp, T.~Hussein, and J.~Ziegler.
\newblock Choice-based preference elicitation for collaborative filtering
  recommender systems.
\newblock In {\em Proceedings of the SIGCHI Conference on Human Factors in
  Computing Systems}, pages 3085--3094. ACM, 2014.

\bibitem{luo2009personalized}
H.~Luo, J.~Fan, D.~A. Keim, and S.~Satoh.
\newblock Personalized news video recommendation.
\newblock In {\em Advances in Multimedia Modeling}, pages 459--471. Springer,
  2009.

\bibitem{6111347}
K.-L. Ma, I.~Liao, J.~Frazier, H.~Hauser, and H.-N. Kostis.
\newblock Scientific storytelling using visualization.
\newblock {\em Computer Graphics and Applications, IEEE}, 32(1):12--19, 2012.

\bibitem{mitchell2011limits}
A.~Mitchell and K.~McGee.
\newblock Limits of rereadability in procedural interactive stories.
\newblock In {\em Proceedings of the SIGCHI Conference on Human Factors in
  Computing Systems}, pages 1939--1948. ACM, 2011.

\bibitem{Pu:2013:UIR:2507157.2507178}
L.~Pu and B.~Faltings.
\newblock Understanding and improving relational matrix factorization in
  recommender systems.
\newblock In {\em Proceedings of the 7th ACM Conference on Recommender
  Systems}, RecSys '13, pages 41--48, 2013.

\bibitem{MovieLens100k}
G.~Research.
\newblock Movielens100k: Movie rating dataset.

\bibitem{1626183}
M.~Riedl and R.~Young.
\newblock From linear story generation to branching story graphs.
\newblock {\em Computer Graphics and Applications, IEEE}, 26(3):23--31, 2006.

\bibitem{Satyanarayan:2014:ANV:2771495.2771532}
A.~Satyanarayan and J.~Heer.
\newblock Authoring narrative visualizations with ellipsis.
\newblock {\em Comput. Graph. Forum}, 33(3):361--370, 2014.

\bibitem{segel2010narrative}
E.~Segel and J.~Heer.
\newblock Narrative visualization: Telling stories with data.
\newblock {\em Visualization and Computer Graphics, IEEE Transactions on},
  16(6):1139--1148, 2010.

\bibitem{spaulding2013design}
E.~Spaulding and H.~Faste.
\newblock Design-driven narrative: using stories to prototype and build
  immersive design worlds.
\newblock In {\em Proceedings of the SIGCHI Conference on Human Factors in
  Computing Systems}, pages 2843--2852. ACM, 2013.

\bibitem{vlachos2012recommendation}
M.~Vlachos and D.~Svonava.
\newblock Recommendation and visualization of similar movies using minimum
  spanning dendrograms.
\newblock {\em Information Visualization}, 2012.

\bibitem{Wang:2016:GTL:2968220.2968242}
Y.~Wang, D.~Liu, H.~Qu, Q.~Luo, and X.~Ma.
\newblock A guided tour of literature review: Facilitating academic paper
  reading with narrative visualization.
\newblock In {\em Proceedings of the International Symposium on Visual
  Information Communication and Interaction}, pages 17--24, 2016.

\bibitem{Wohlfart:2007:STP:2384179.2384194}
M.~Wohlfart and H.~Hauser.
\newblock Story telling for presentation in volume visualization.
\newblock In {\em Proceedings of the 9th Joint Eurographics / IEEE VGTC
  Conference on Visualization}, EUROVIS'07, pages 91--98, 2007.

\bibitem{wojtkowski2002storytelling}
W.~Wojtkowski and W.~G. Wojtkowski.
\newblock Storytelling: its role in information visualization.
\newblock In {\em European Systems Science Congress}. Citeseer, 2002.

\bibitem{CGF:CGF1816}
L.~Yu, A.~Lu, W.~Ribarsky, and W.~Chen.
\newblock Automatic animation for time-varying data visualization.
\newblock {\em Computer Graphics Forum}, 29(7):2271--2280, 2010.

\end{thebibliography}
\end{document}